\documentclass[a4paper,fleqn]{cas-sc}
\usepackage[numbers]{natbib}

\usepackage{amssymb}
\usepackage{amsmath}
\usepackage{mathtools}
\usepackage{booktabs}
\usepackage{caption}
\usepackage{subcaption}
\usepackage{graphicx}
\usepackage{placeins}
\usepackage{chngcntr}
\counterwithin{figure}{section}
\usepackage{cuted}
\usepackage[nameinlink,capitalize]{cleveref}
\usepackage{hyperref}
\usepackage{float}
\usepackage{eufrak}

\graphicspath{{figures/}}

\newcommand{\SU}{\mathrm{SU}}
\newcommand{\ket}[1]{\left|#1\right\rangle}
\newcommand{\bra}[1]{\left\langle#1\right|}

\begin{document}
\let\WriteBookmarks\relax
\def\floatpagepagefraction{1}
\def\textpagefraction{.001}

%% ---------- Short (running-head) title and authors ----------
\shorttitle{Multi-channel collective dissipation via $\SU(4)$}
\shortauthors{M. Lutsukh et al.}

%% ---------- Title ----------
\title[mode=title]{Multi-channel collective dissipation via the symmetric
irreducible representation of \texorpdfstring{$\SU(4)$}{SU(4)}}

\author[1,2]{Mend-Amar Lutsukh}
\author[1,2]{Munkh-Uchral Bazarsan}
\author[2]{Tuguldur Begzjav}
\author[3,4,5]{Gombojav O. Ariunbold\corref{cor1}}
\ead{ag2372@msstate.edu}
\cortext[cor1]{Corresponding author.}

%% ---------- Affiliations ----------

\affiliation[1]{organization={Institute of Physics and Technology,
            Mongolian Academy of Sciences},
            city={Ulaanbaatar},
            postcode={13330},
            country={Mongolia}}
        
\affiliation[2]{organization={Department of Physics,
            National University of Mongolia},
            city={Ulaanbaatar},
            postcode={14200},
            country={Mongolia}}

\affiliation[3]{organization={Department of Physics and Astronomy,
            Mississippi State University},
            city={Mississippi State},
            postcode={39762},
            state={MS},
            country={USA}}

\affiliation[4]{organization={Physics Department, Wesleyan University},
            city={Middletown},
            postcode={06459},
            state={CT},
            country={USA}}

\affiliation[5]{organization={Department of Physics and Astronomy,
            Texas A\&M University},
            city={College Station},
            state={TX},
            country={USA}}

\begin{abstract}
We specialize Agarwal's multi-level collective spontaneous-emission
formalism to the four-level case by formulating it in the fully symmetric
$\SU(4)$ representation of $N$ identical atoms. In the irreducible
representation $(N,0,0)$, the occupation-number basis forms a tetrahedral
weight lattice on which the six embedded $\mathfrak{su}(2)$ transition
subalgebras act as ladder operators. From these algebraic factors we
obtain a compact Pauli-type population-rate equation and a closed-form
expression for the total emitted intensity that apply to any combination
of open dipole channels. The formalism is then specialized to the seven
dipole-allowed four-level topologies---tripod, inverted tripod, Y,
inverted Y, double-$\Lambda$, closed cascade, and diamond---and the
resulting rate equations are solved numerically for atom numbers up to
$N=50$. In every case the emitted intensity develops a delayed
cooperative burst whose peak height obeys a power law
$I_{\mathrm{peak}}=aN^{p}$ with topology-dependent parameters
$(a,p)$; the fitted exponents lie in the range $1.81\lesssim p\lesssim 1.92$,
indicating a superlinear. The $\SU(4)$ tetrahedral flow and the
seven configuration-dependent transients together provide a unified
geometric picture of multi-channel collective dissipation in four-level
atomic ensembles.
\end{abstract}

\begin{highlights}
\item An $\SU(4)$-symmetric master equation is derived that
unifies all seven dipole-allowed four-level topologies in a single
Pauli-type rate equation on the tetrahedral weight lattice.
\item Closed-form expressions are obtained for the collective transition
matrix elements and for the total emitted intensity in the fully
symmetric $(N,0,0)$ representation.
\item The rate equation is solved numerically for all seven topologies, and the resulting probability flow is
visualized directly on the $\SU(4)$ tetrahedron.
\item Peak intensities obey a topology-dependent power law
$I_{\mathrm{peak}}=aN^{p}$ with fitted exponents
$1.81\lesssim p\lesssim 1.92$ over the range $N=20$--$40$.
\item The formalism provides a common four-level extension of our recent
three-level $\SU(3)$ trilogy (cascade, $\Lambda$, and V-type
configurations) and a natural starting point for adding coherences,
propagation, and multi-photon correlations.
\end{highlights}

\begin{keywords}
Dicke superradiance \sep Collective spontaneous emission \sep Multi-mode superradiance \sep $\SU(4)$ symmetry \sep
Four-level atoms \sep Pauli-type master equation \sep Multi-channel collective dissipation
\end{keywords}

\maketitle

\section{Introduction}
\label{sec:intro}

Cooperative spontaneous emission from an ensemble of identical two-level
atoms---\emph{Dicke superradiance}---is one of the few genuinely many-body
effects of quantum electrodynamics that admits an analytically tractable
description \cite{dicke1954}. In his 1954 paper, Dicke showed that a small
sample of $N$ two-level atoms confined to a volume much smaller than the emission
wavelength bursts a delayed, intense pulse whose
peak power scales as $N^{2}$ rather than as the incoherent $N$ result. The
symmetric subspace of the $N$ atoms is spanned by only $N+1$ collective states
$\ket{J,M}$, and the collective ladder matrix element
$\sqrt{(J-M)(J+M+1)}$ produces the Bose-like enhancement responsible for the
$N^{2}$ scaling. The quantum statistical theory of this cooperative emission
was subsequently developed by Rehler and Eberly \cite{rehler1971}, by
Bonifacio, Schwendimann and Haake \cite{bonifacio1971a,bonifacio1971b}, by
Bonifacio and Lugiato \cite{bonifacio1975}, by Degiorgio \cite{degiorgio1971},
and by MacGillivray and Feld for the extended, optically thick case
\cite{macgillivray1976}, and was subsequently reviewed in a widely read essay
by Gross and Haroche \cite{gross1982} and a monograph by Benedict et al.\
\cite{benedict1996}. Exact analytical solutions of the two-level master
equation \cite{lee1977a,lee1977b} and the associated statistics of tipping
angles and delay times \cite{lee1982} completed the theoretical picture of the
two-level case. Experimentally, the first observation of Dicke superradiance
was reported by Skribanowitz et al.\ in optically pumped HF gas
\cite{skribanowitz1973}, followed by measurements of macroscopic quantum
fluctuations in the emission delay time \cite{vrehen1980,vrehen1981}. Later
work explored single-photon Dicke superradiance and the collective Lamb shift
\cite{svidzinsky2008,rohlsberger2010,scully2009}.

Interest in cooperative emission has resurged in the last fifteen years
because the phenomenon has now been demonstrated in a remarkable variety of
condensed-phase and cold-atom platforms. Room-temperature superradiance from
single diamond nanocrystals \cite{bradac2017}, superfluorescence from
lead-halide perovskite quantum-dot superlattices \cite{raino2018}, and
high-temperature and even room-temperature superfluorescence from
methyl-ammonium lead iodide and hybrid perovskites
\cite{findik2021,biliroglu2022,huang2022} have shown that macroscopic quantum
coherence can survive in solid-state materials under ambient conditions.
Steady-state superradiant lasers with sub-photon intracavity power
\cite{bohnet2012} and superradiant emission on the mHz-linewidth strontium
clock transition \cite{norcia2016} have opened routes to next-generation
optical clocks and mHz-linewidth lasers. The persistent theoretical interest
in these systems is reflected by broad reviews for solids \cite{cong2016} and
for generation-and-implementation aspects in laser physics
\cite{kocharovsky2017,ariunbold2022}, and by experiments in atomic vapors
\cite{ariunbold2010,ariunbold2012ol,thompson2014,ariunbold2014apl,ariunbold2022csbeats},
in cryogenically pumped rare-earth-doped solids
\cite{braggio2020,chiossi2021}, and in the extreme-ultraviolet regime.

The three-level generalization of Dicke's model marks the first level of
multi-mode complexity and was already anticipated in the pioneering works of
Agarwal \cite{agarwal1970,agarwal1973}, of Agarwal and Trivedi
\cite{agarwal1976}, and of Cho, Kurnit and Gilmore \cite{cho1973,gilmore1973}.
Formally, the symmetric subspace of $N$ identical three-level atoms is
spanned by the fully symmetric irreducible representation of $\SU(3)$, whose
weight lattice is a two-dimensional triangular subdivision. Exact numerical
solutions of the corresponding two-mode rate equations, however, remained
essentially unexplored for almost five decades. Recent work by our group has
returned to this problem for the three canonical topologies of a three-level
atom. In the ladder (cascade) configuration, the cascade superradiance model
of Ref.~\cite{ariunbold2022csr} extended Dicke's simple two-level formula to
the two-mode cooperative emission, with a fully time-dependent treatment of
the average delay, its quantum-mechanical fluctuations, and the correlations
between the two modes. For the $\Lambda$-type three-level atom, in which two
optical modes share a common upper level, we obtained the first exact
numerical solutions of the two-mode rate equations and
identified a novel mode-selective control regime in which the properties of
one Dicke-like burst can be tuned through the parameters of the other
\cite{ariunbold2025lambda}. Most recently, the $V$-type three-level atom was shown to
support a two-pulse ``superradiant synthesis'' by which two decoupled
sub-ensembles are dynamically merged into a single macroscopic collective
state \cite{ariunbold2026vee}. These three configurations---ladder, $\Lambda$
and $V$---together exhaust the dipole-allowed three-level topologies and are
the three-level $\SU(3)$ analogues of the seven four-level $\SU(4)$
configurations analyzed in the present paper.

The natural next step---the four-level, $d=4$ generalization---is important
not only as a mathematical unification but for several concrete physical
reasons. First effect of collective dissipation due to the open quantum system is of great interest as four-level configurations underlie most schemes for coherent
population trapping, electromagnetically induced transparency, slow light,
and photon-pair generation by four-wave mixing
\cite{fleischhauer2005,mazets2005,niu2002,shu2016,unanyan1998,lukin1997},
including the double-$\Lambda$ EIT biphoton source \cite{cho2016,tey2008}.
Second, recent experiments on alkali vapors (Rb, Cs, Na) explicitly probe
two-photon and multi-step transitions that link four internal levels
\cite{ariunbold2010,thompson2014,ariunbold2022csbeats}. Third, the additional Hilbert-space
degeneracy of a four-level SU(4)-symmetric ensemble endows the emitted
biphotons and multiphotons with a richer entanglement structure than in the
two- or three-level case \cite{pandey2025,garziano2016} and it is important to study how this structure behaves in the overdamped regime due to the collective dissipation. Despite this
convergence of interest, a common $\SU(4)$-symmetric formulation that treats
all dipole-allowed four-level configurations on the same footing has, to our
knowledge, not been written down.

In this paper, we partially close this gap. Starting from the multi-level dissipative
master equation of Agarwal \cite{agarwal1970,agarwal1973,agarwal1976}, we
evaluate how the collective raising and lowering operators of $\mathfrak{su}(4)$
act on the fully symmetric occupation-number basis
$\ket{q_1,q_2,q_3,q_4}$, where $q_i$ counts the atoms in the $i$-th level
($i=1,2,3,4$) and $q_1+q_2+q_3+q_4=N$. \Cref{sec:su4-representation} develops
the $\SU(4)$ representation framework: it introduces the fundamental and
fully symmetric representations, the Lie algebra $\mathfrak{su}(4)$, and its
six embedded $\mathfrak{su}(2)$ transition subalgebras, and writes the
collective transition operators $A_{nm}$ explicitly in the symmetric basis.
\Cref{sec:master-equation} uses those matrix elements to obtain a Pauli-type
rate equation for the joint occupation-number distribution
$p(q_1,q_2,q_3,q_4,t)$, together with a compact closed-form expression for
the total emitted intensity $I(t)$; it also derives the exact energy-balance
identity $d\langle E\rangle/dt=-I(t)$ that later serves as a stringent
numerical benchmark. \Cref{sec:configs} specializes the master equation to
the seven dipole-allowed four-level topologies selected by parity, namely
the tripod, inverted tripod, Y, inverted-Y, double-$\Lambda$, closed-cascade
and diamond schemes. \Cref{sec:numerics} solves the resulting rate equation
numerically for atom numbers up to $N=50$: it visualizes the probability
flow on the $\SU(4)$ tetrahedron, examines the emitted-intensity transients,
extracts the scaling of the superradiant peak
$I_{\mathrm{peak}}(N)=aN^{p}$ across all seven configurations, and benchmarks
probability and energy conservation. \Cref{sec:discussion} places the
formalism in the context of current experimental platforms;
\Cref{sec:conclusion} concludes and outlines directions for future work.

\section{Symmetric \texorpdfstring{$\SU(4)$}{SU(4)} representation of four-level atoms}
\label{sec:su4-representation}

\subsection{Four-level atoms and the fundamental representation of \texorpdfstring{$\SU(4)$}{SU(4)}}

A single four-level atom lives in the four-dimensional internal Hilbert
space $\mathcal{H}_{1}=\mathrm{span}\{\ket{1},\ket{2},\ket{3},\ket{4}\}
\cong\mathbb{C}^{4}$, with orthonormal basis
$\{\ket{1},\ket{2},\ket{3},\ket{4}\}$ ordered by energy
$E_{1}>E_{2}>E_{3}>E_{4}$. The natural unitary transformations of this
internal state space form the group $\mathrm{U}(4)$, and since an
overall phase is unobservable the relevant group is the special
unitary group $\SU(4)$. Its Lie algebra $\mathfrak{su}(4)$ has
dimension $4^{2}-1=15$ and, in the physics convention, is spanned by
Hermitian traceless generators $\{T_{a}\}_{a=1}^{15}$ satisfying
\begin{equation}
[T_{a},T_{b}]= i f_{abc}T_{c},
\end{equation}
with $f_{abc}$ the structure constants. A basis better suited to the
optics problem is provided by the fifteen independent traceless
combinations of the transition operators
\begin{equation}
A_{nm}=\ket{n}\bra{m},\qquad n,m=1,2,3,4, \qquad \sum_{i=n}^4=\ket{n}\bra{n}=\mathbb{I}
\label{eq:transition-operators}
\end{equation}
which for $n\neq m$ transfer population from level $m$ to level $n$.
The transition operators satisfy the commutation relation
\begin{equation}
[A_{nm},A_{kl}]
=
\delta_{mk}A_{nl}
-
\delta_{nl}A_{km}.
\label{eq:su4-commutation}
\end{equation}
This relation provides the algebraic basis for constructing the collective transition operators.

\begin{figure*}[h]
\centering
\includegraphics[width=0.95\textwidth]{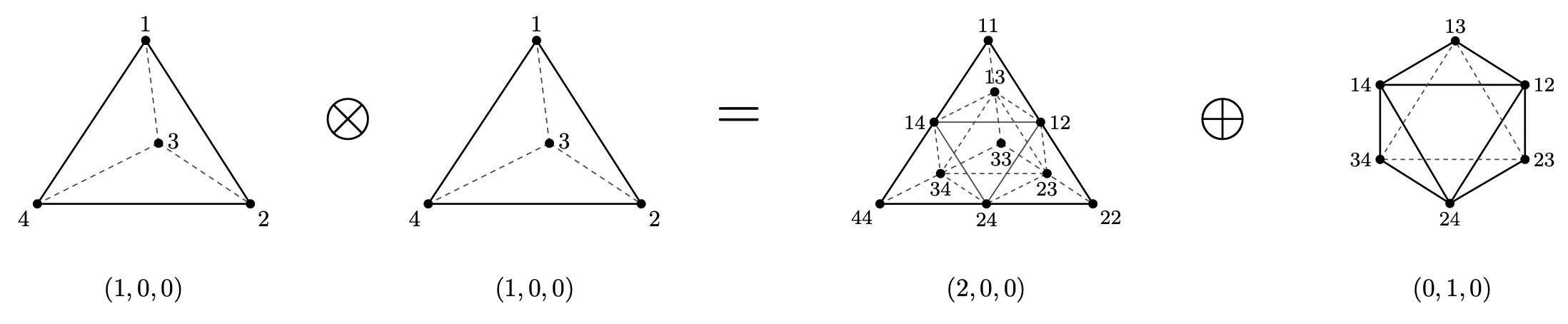}
\caption{
Tensor-product decomposition of two fundamental $\SU(4)$ representations, $(1,0,0)\otimes(1,0,0)=(2,0,0)\oplus(0,1,0)$. The
four vertices labeled $1,2,3,4$ represent the one-particle basis states
$\{\ket{1},\ket{2},\ket{3},\ket{4}\}$ of the fundamental representation
$(1,0,0)$. The representation $(2,0,0)$ corresponds to the fully symmetric subspace: labels $11,22,33,44$ denote two particles in the same internal state, while labels \(ij\) with \(i<j\) denote symmetrized states \((\ket{i}\ket{j}+\ket{j}\ket{i})/\sqrt{2}\). The representation $(0,1,0)$ corresponds to the antisymmetric subspace, with labels $ij$ denoting $(\ket{i}\ket{j}-\ket{j}\ket{i})/\sqrt{2}$. Solid lines show visible edges of the projected three-dimensional diagram, and dashed lines show hidden or interior edges.
}
\label{fig:n2-weight}
\end{figure*}

\Cref{fig:n2-weight} illustrates the tensor-product decomposition of two
fundamental $\SU(4)$ representations, corresponding to the \(N=2\) case. Since $\mathfrak{su}(4)$ has rank three, its weight space is three-dimensional. The four weights of the fundamental representation $(1,0,0)$ can be placed at the
vertices of a tetrahedron. In this geometric picture, the vertices labeled
$1,2,3,4$ represent the one-particle states
$\ket{1},\ket{2},\ket{3},\ket{4}$. A transition operator $A_{nm}$ moves a state
from the weight associated with level $m$ to the weight associated with level
$n$.
For two four-level atoms, the composite Hilbert space is
$\mathcal{H}_1\otimes\mathcal{H}_1$ with dimension $4\times4=16$. In $\SU(4)$ representation notation this product
is $(1,0,0)\otimes(1,0,0).$ This tensor-product representation is not irreducible. The reason is that the
action of $\SU(4)$ is the same on both particles and therefore commutes with
the exchange of the two particles. Consequently, the two-particle space splits
into two invariant parts: a symmetric subspace and an antisymmetric subspace.
This gives the irreducible decomposition
\begin{equation}
(1,0,0)\otimes(1,0,0)
=
(2,0,0)\oplus(0,1,0).
\label{eq:su4-two-particle-decomposition}
\end{equation}
The representation $(2,0,0)$ is the fully symmetric two-particle
representation and has dimension $10$. The representation $(0,1,0)$ is the
antisymmetric two-particle representation and has dimension $6$.

\subsection{Fully symmetric \texorpdfstring{$N$}{N}-atom representation}

In the collective-emission problem considered here, the relevant many-body
sector is the permutation-symmetric subspace of $N$ identical atoms. This
subspace is spanned by occupation-number states $\ket{q_1,q_2,q_3,q_4}$.
This symmetric subspace carries the fully symmetric irreducible representation
of $\SU(4)$ with Dynkin label $(N,0,0).$
The dimension of this representation is the number of ways to distribute $N$
identical atoms among four internal levels:
\begin{equation}
D_N
=
\binom{N+3}{3}
=
\frac{(N+1)(N+2)(N+3)}{6}.
\label{eq:su4-symmetric-dimension}
\end{equation}

The corresponding weight of the state $\ket{q_1,q_2,q_3,q_4}$ is
\begin{equation}
\boldsymbol{\Lambda}
=
q_1\boldsymbol{\lambda}_1
+
q_2\boldsymbol{\lambda}_2
+
q_3\boldsymbol{\lambda}_3
+
q_4\boldsymbol{\lambda}_4.
\end{equation}
Here, \(\boldsymbol{\lambda}_i\) denotes the weight vector associated with the basis state \(\ket{i}\). Because the occupation numbers obey
$q_1+q_2+q_3+q_4=N$, all allowed weights lie on a finite tetrahedral lattice.
Thus the general $N$-particle weight diagram is a tetrahedron subdivided into
$N$ steps along each edge. The lattice sites are in one-to-one correspondence
with the symmetric occupation states $\ket{q_1,q_2,q_3,q_4}$.

This tetrahedral geometry is useful because the collective dynamics can be
viewed as probability flow between neighboring lattice sites. A transition
operator $A_{nm}$ changes the occupation numbers by
\begin{equation}
q_n\rightarrow q_n+1,
\qquad
q_m\rightarrow q_m-1,
\end{equation}
and therefore moves the system along an edge direction of the tetrahedral
weight lattice.

\subsection{\texorpdfstring{$\mathfrak{su}(4)$}{su(4)} subalgebras and transition multiplets}
\label{subsec:submultiplets}

Each pair of levels $(a,b)$ defines an embedded $\mathfrak{su}(2)$ subalgebra
inside $\mathfrak{su}(4)$. For a fixed pair $(a,b)$, define
\begin{equation}
J_+^{(ab)}=A_{ab},
\qquad
J_-^{(ab)}=A_{ba},
\qquad
J_z^{(ab)}=\frac{1}{2}(A_{aa}-A_{bb}).
\end{equation}
Using Eq.~\eqref{eq:su4-commutation}, these operators satisfy
\begin{equation}
[J_z^{(ab)},J_\pm^{(ab)}]=\pm J_\pm^{(ab)},
\qquad
[J_+^{(ab)},J_-^{(ab)}]=2J_z^{(ab)}.
\end{equation}
Thus, every pair of levels behaves as a two-level subsystem embedded in the
four-level atom. Since four levels contain six unordered pairs, $\mathfrak{su}(4)$
contains six such $\mathfrak{su}(2)$ transition subalgebras.

In the symmetric occupation basis, the collective transition operators are
written in Schwinger form,
\begin{equation}
A_{nm}=a_n^\dagger a_m.
\end{equation}
Here $a_m$ removes one atom from level $m$, while $a_n^\dagger$ places one atom
in level $n$. Their action on a symmetric occupation state is
\begin{equation}
A_{nm}\ket{q_1,q_2,q_3,q_4}
=
\sqrt{q_m(q_n+1)}
\ket{\ldots,q_n+1,\ldots,q_m-1,\ldots}.
\label{eq:anm-action}
\end{equation}
This equation is the key algebraic result of the section. The operator $A_{nm}$
transfers one atom from level $m$ to level $n$, and the square-root factor is
the collective matrix element associated with the symmetric representation.

As an example, the subalgebra connecting levels $1$ and $3$ is generated by
$A_{13}$, $A_{31}$, and $(A_{11}-A_{33})/2$. Its ladder action is
\begin{align}
A_{13}\ket{q_1,q_2,q_3,q_4}
&=
\sqrt{q_3(q_1+1)}
\ket{q_1+1,q_2,q_3-1,q_4},
\\
A_{31}\ket{q_1,q_2,q_3,q_4}
&=
\sqrt{q_1(q_3+1)}
\ket{q_1-1,q_2,q_3+1,q_4}.
\end{align}
These operators raise and lower
\begin{equation}
m_{13}=\frac{q_1-q_3}{2}
\end{equation}
at fixed
\begin{equation}
j_{13}=\frac{q_1+q_3}{2}
=
\frac{N-q_2-q_4}{2}.
\end{equation}

The six embedded $\mathfrak{su}(2)$ multiplets are summarized in
Table~\ref{tab:submultiplets}. Each multiplet corresponds to motion along one
family of straight lines in the tetrahedral weight lattice.

\begin{table}[t]
\centering
\caption{
Six embedded $\mathfrak{su}(2)$ subalgebras of $\mathfrak{su}(4)$ in the
fully symmetric $N$-atom representation. For each pair of levels $(a,b)$,
$A_+=A_{ab}$ and $A_-=A_{ba}$ raise and lower
$m=(q_a-q_b)/2$ at fixed $j=(q_a+q_b)/2$. The equivalent expressions for $j$
below use the constraint $q_1+q_2+q_3+q_4=N$.
}
\label{tab:submultiplets}
\begin{tabular}{ccccc}
\toprule
Multiplet & Levels $(a,b)$ & $A_+$ & $A_-$ & $(j,m)$\\
\midrule
$T$ & $(1,2)$ & $A_{12}$ & $A_{21}$ &
$\left(\frac{N-q_3-q_4}{2},\frac{q_1-q_2}{2}\right)$\\[3pt]
$V$ & $(1,3)$ & $A_{13}$ & $A_{31}$ &
$\left(\frac{N-q_2-q_4}{2},\frac{q_1-q_3}{2}\right)$\\[3pt]
$W$ & $(1,4)$ & $A_{14}$ & $A_{41}$ &
$\left(\frac{N-q_2-q_3}{2},\frac{q_1-q_4}{2}\right)$\\[3pt]
$U$ & $(2,3)$ & $A_{23}$ & $A_{32}$ &
$\left(\frac{N-q_1-q_4}{2},\frac{q_2-q_3}{2}\right)$\\[3pt]
$X$ & $(2,4)$ & $A_{24}$ & $A_{42}$ &
$\left(\frac{N-q_1-q_3}{2},\frac{q_2-q_4}{2}\right)$\\[3pt]
$Z$ & $(3,4)$ & $A_{34}$ & $A_{43}$ &
$\left(\frac{N-q_1-q_2}{2},\frac{q_3-q_4}{2}\right)$\\
\bottomrule
\end{tabular}
\end{table}

Equation~\eqref{eq:anm-action} is the $\SU(4)$ analogue of the familiar
$\mathfrak{su}(2)$ ladder matrix element
\begin{equation}
\sqrt{(j-m)(j+m+1)}.
\end{equation}
It is this representation-theoretic matrix element that produces the
collective enhancement factors in the population-rate equation and in the
emitted intensity derived in the following section.

The above construction has been purely algebraic. Once the collective
transition operators $A_{nm}$ and their matrix elements in the symmetric
basis $\ket{q_1,q_2,q_3,q_4}$ are known, the many-body dissipative
dynamics of $N$ identical four-level atoms are fully determined by
substituting Eq.~\eqref{eq:anm-action} into the standard multi-level
master equation. This substitution is carried out in the following
section, where the resulting Pauli-type equation is shown to close on
the diagonal populations $p(q_1,q_2,q_3,q_4,t)$ and to yield a compact
expression for the total emitted intensity.

\section{Population master equation on the \texorpdfstring{$\mathrm{SU}(4)$}{SU(4)} tetrahedral lattice}
\label{sec:master-equation}

Building on the algebraic action of the collective ladder operators
$A_{nm}$ derived in \Cref{subsec:submultiplets}, we now translate the
Lindblad dynamics of $N$ identical four-level atoms into a rate equation
for the joint occupation-number distribution
$p(q_1,q_2,q_3,q_4,t)$. The master equation is written in Lindblad
form as
\begin{equation}
\frac{\partial \rho}{\partial t}
=
\sum_{n>m}
\gamma_{nm}
\left(
2A_{nm}\rho A_{mn}
-
A_{mn}A_{nm}\rho
-
\rho A_{mn}A_{nm}
\right),
\label{eq:lindblad-su4}
\end{equation}
where $\gamma_{nm}$ is the single-atom spontaneous-emission rate for the
transition $m\rightarrow n$. Equivalently,
\begin{equation}
\gamma_{nm}
=
\frac{\omega_{nm}^{3}|d_{nm}|^2}
{3\pi\epsilon_0\hbar c^3},
\qquad
\omega_{nm}
=
\frac{E_m-E_n}{\hbar}>0.
\label{eq:gamma-rate}
\end{equation}
The notation $n>m$ is therefore an index convention for downward transitions,
not an energy ordering.

We are interested in the population dynamics on the symmetric weight lattice.
Define
\begin{equation}
p(\mathbf{q},t)
\equiv
p(q_1,q_2,q_3,q_4,t)
=
\bra{q_1,q_2,q_3,q_4}
\rho(t)
\ket{q_1,q_2,q_3,q_4}.
\end{equation}
Using the ladder action
\begin{equation}
A_{nm}\ket{\mathbf{q}}
=
\sqrt{q_m(q_n+1)}
\ket{\mathbf{q}+\mathbf{e}_n-\mathbf{e}_m},
\end{equation}
where $\mathbf{e}_i$ is the unit vector in the $q_i$ direction, the diagonal
part of Eq.~\eqref{eq:lindblad-su4} closes on the populations alone. For each
open channel $m\rightarrow n$, probability leaves the state $\mathbf{q}$ at the
collective rate
\begin{equation}
2\gamma_{nm}q_m(q_n+1),
\end{equation}
and enters $\mathbf{q}$ from the parent state
$\mathbf{q}-\mathbf{e}_n+\mathbf{e}_m$. The resulting Pauli-type rate equation is
\begin{equation}
\frac{\partial p(\mathbf{q},t)}{\partial t}
=
\sum_{n>m}
2\gamma_{nm}
\left[
q_n(q_m+1)
p(\mathbf{q}-\mathbf{e}_n+\mathbf{e}_m,t)
-
q_m(q_n+1)
p(\mathbf{q},t)
\right].
\label{eq:pauli-rate-compact}
\end{equation}
Terms containing negative occupation numbers are omitted. This equation is the
population-space form of the collective master equation on the symmetric
$\mathrm{SU}(4)$ tetrahedron.

For clarity, we write Eq.~\eqref{eq:pauli-rate-compact} explicitly for the six
possible downward channels:
\begin{equation}
\small
\begin{aligned}
\frac{\partial p}{\partial t}
={}&
2\gamma_{21}
\Big[
q_2(q_1+1)
p(q_1+1,q_2-1,q_3,q_4)
-
q_1(q_2+1)p
\Big]
\\
&+
2\gamma_{31}
\Big[
q_3(q_1+1)
p(q_1+1,q_2,q_3-1,q_4)
-
q_1(q_3+1)p
\Big]
\\
&+
2\gamma_{41}
\Big[
q_4(q_1+1)
p(q_1+1,q_2,q_3,q_4-1)
-
q_1(q_4+1)p
\Big]
\\
&+
2\gamma_{32}
\Big[
q_3(q_2+1)
p(q_1,q_2+1,q_3-1,q_4)
-
q_2(q_3+1)p
\Big]
\\
&+
2\gamma_{42}
\Big[
q_4(q_2+1)
p(q_1,q_2+1,q_3,q_4-1)
-
q_2(q_4+1)p
\Big]
\\
&+
2\gamma_{43}
\Big[
q_4(q_3+1)
p(q_1,q_2,q_3+1,q_4-1)
-
q_3(q_4+1)p
\Big].
\end{aligned}
\label{eq:all-six-channels}
\end{equation}
Here $p$ without an argument denotes
$p(q_1,q_2,q_3,q_4,t)$. Equation~\eqref{eq:all-six-channels} is the explicit
four-level $\mathrm{SU}(4)$ probability-rate equation. It is not necessary to
derive a separate master equation for each four-level topology. A specific
level scheme is obtained simply by setting the forbidden decay rates
$\gamma_{nm}$ to zero.

Probability conservation follows directly from the gain-loss structure of
Eq.~\eqref{eq:pauli-rate-compact}. Summing over all lattice points
$\mathbf{q}$ gives
\begin{equation}
\frac{d}{dt}\sum_{\mathbf{q}}p(\mathbf{q},t)=0,
\end{equation}
because every loss term from one lattice site appears as a gain term at the
neighboring site reached by the corresponding transition.

The emitted intensity is obtained by weighting every jump by the photon energy
$\hbar\omega_{nm}$. Therefore,
\begin{equation}
I(t)
=
2
\sum_{n>m}
\hbar\omega_{nm}\gamma_{nm}
\sum_{\mathbf{q}}
p(\mathbf{q},t)
q_m(q_n+1).
\label{eq:intensity}
\end{equation}
We set $\hbar=1$. The collective factor $q_m(q_n+1)$ appearing in
Eq.~\eqref{eq:intensity} has a direct physical interpretation: $q_m$
counts how many atoms are available in the emitting upper level $m$,
while $q_n+1$ is the bosonic enhancement associated with the symmetric
final state after one atom has joined level $n$. This is the four-level
$\SU(4)$ generalization of the Dicke $(J+M)(J-M+1)$ factor familiar
from a two-level ensemble.

Equation~\eqref{eq:intensity} also makes the geometrical origin of the
superradiant burst transparent. During the evolution, probability flows through
the tetrahedral weight lattice. Whenever a decay channel connects a
macroscopically occupied upper level to a lower level that is also becoming
macroscopically occupied, the product $q_m(q_n+1)$ becomes of order $N^2$.
Consequently, the emitted intensity can grow superlinearly with atom number,
producing the delayed collective peak characteristic of superradiance.

For later use, it is useful to record the corresponding mean energy balance.
With
\begin{equation}
\langle E(t)\rangle
=
\sum_{\mathbf{q}}
p(\mathbf{q},t)
\sum_{i=1}^{4}q_iE_i,
\end{equation}
the rate equation gives
\begin{equation}
\frac{d\langle E(t)\rangle}{dt}
=
-
I(t).
\label{eq:energy-balance}
\end{equation}
Thus the emitted intensity is exactly the rate at which internal atomic energy
is lost. This identity provides a useful numerical check on the implementation
of the $\mathrm{SU}(4)$ population dynamics.

\section{Seven dipole-allowed four-level configurations}
\label{sec:configs}

\begin{figure*}[t]
\centering
\includegraphics[width=0.90\linewidth]{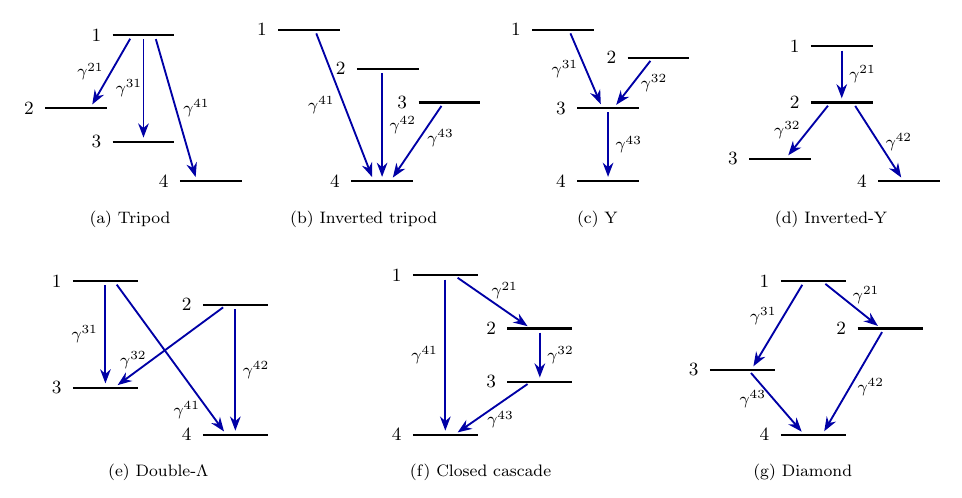}
\caption{The seven dipole-allowed four-level interaction topologies selected by
parity: tripod, inverted tripod, Y, inverted Y, double-$\Lambda$, closed
cascade, and diamond. Arrows mark the open decay channels $\gamma^{nm}$ in each
configuration.}
\label{fig:seven-configs}
\end{figure*}

The general rate equation~\eqref{eq:pauli-rate-compact} contains, in
principle, contributions from all six possible downward channels between
the four internal levels. In a physical atom, however, only a small
subset of these channels is optically active: the parity selection rule
of the electric-dipole interaction $H_{\mathrm{int}}=-\hat{\mathbf{d}}
\cdot\mathbf{E}$ requires that the two connected states have opposite
parity, so no more than four of the six rates can simultaneously be
nonzero. Enumerating the topologically distinct sign patterns of the
six-channel rate matrix under this constraint gives exactly seven
inequivalent four-level configurations---the tripod, inverted tripod,
Y, inverted-Y, double-$\Lambda$, closed-cascade, and diamond schemes
of \cref{fig:seven-configs}. These seven topologies cover the great majority of dipole-coupled
four-level systems reported in the literature:
tripod schemes have been realized in metastable Ne and $^{87}\mathrm{Rb}$
\cite{mazets2005,unanyan1998}, diamond configurations in Rb vapors for
double-EIT and four-wave mixing \cite{niu2002}, closed-cascade
emission in cold-atom ensembles \cite{wang2015} and rare-earth-doped
solids \cite{chiossi2021}, double-$\Lambda$ biphoton sources in EIT
systems \cite{cho2016,shu2016}, and inverted-Y EIT phenomena
\cite{lukin1997}. A central appeal of the $\SU(4)$-symmetric formulation
of \Cref{sec:master-equation} is that all seven topologies are described
by the same population master equation
\eqref{eq:pauli-rate-compact}: what changes from one configuration to
the next is only the pattern of open decay channels, i.e.\ the sign
mask of the rates $\gamma_{nm}$ collected in
\cref{tab:seven-configs}.

For each configuration, \cref{eq:all-six-channels} collapses to a master
equation by setting all decay constants not listed in
\cref{tab:seven-configs} to zero. Thus, the table contains the complete
configuration-dependent input to the population dynamics.

\begin{table}[t]
\centering
\normalfont\rmfamily\footnotesize
\caption{Seven four-level configurations and their nonzero spontaneous-emission
decay constants. All decay constants not shown are set to zero.}
\label{tab:seven-configs}
\setlength{\tabcolsep}{3.5pt}
\renewcommand{\arraystretch}{1.15}
\begin{tabular*}{\columnwidth}{@{\extracolsep{\fill}}ll@{}}
\toprule
Configuration & Nonzero decay constants \\
\midrule
Tripod & $\gamma_{21},\gamma_{31},\gamma_{41}$ \\
Inverted tripod & $\gamma_{41},\gamma_{42},\gamma_{43}$ \\
Y & $\gamma_{31},\gamma_{32},\gamma_{43}$ \\
Inverted Y & $\gamma_{21},\gamma_{32},\gamma_{42}$ \\
Double-$\Lambda$ & $\gamma_{31},\gamma_{41},\gamma_{32},\gamma_{42}$ \\
Closed cascade & $\gamma_{21},\gamma_{32},\gamma_{43},\gamma_{41}$ \\
Diamond & $\gamma_{21},\gamma_{31},\gamma_{42},\gamma_{43}$ \\
\bottomrule
\end{tabular*}
\end{table}

As a representative example, consider the tripod configuration. The only
nonzero decay constants are $\gamma_{21},\; \gamma_{31},\;\gamma_{41},$ corresponding to the transitions \(1\to2\), \(1\to3\), and \(1\to4\),
respectively. Hence the tripod master equation can be written as
\begin{equation}
\small
\begin{aligned}
\frac{\partial p}{\partial t}
={}&
2\gamma_{21}
\Big[
q_2(q_1+1)
p(q_1+1,q_2-1,q_3,q_4)
-
q_1(q_2+1)p
\Big]
\\
&+
2\gamma_{31}
\Big[
q_3(q_1+1)
p(q_1+1,q_2,q_3-1,q_4)
-
q_1(q_3+1)p
\Big]
\\
&+
2\gamma_{41}
\Big[
q_4(q_1+1)
p(q_1+1,q_2,q_3,q_4-1)
-
q_1(q_4+1)p
\Big],
\end{aligned}
\label{eq:tripod-master}
\end{equation}
where \(p\equiv p(q_1,q_2,q_3,q_4,t)\). Terms with negative occupation numbers
are omitted. The first term describes emission from level \(1\) to level \(2\),
the second from level \(1\) to level \(3\), and the third from level \(1\) to
level \(4\).

The tripod, inverted tripod, Y, and inverted-Y configurations contain three
nonzero decay constants, whereas the double-$\Lambda$, closed-cascade, and
diamond configurations contain four. In every case, no separate master
equation has to be derived: the appropriate population-rate equation
follows directly from \cref{eq:all-six-channels} after the forbidden
transitions are removed by setting their decay constants to zero. This
common origin will be used in the next section, where the same numerical
solver is applied to all seven configurations and their emission
transients are compared side by side.

\section{Numerical results}
\label{sec:numerics}

Having reduced the collective dynamics of every four-level topology to a
single Pauli-type rate equation with a configuration-dependent decay
pattern, we now solve that equation numerically and use the resulting
solutions to compare the seven configurations quantitatively. The
symmetric $\SU(4)$ subspace has finite but polynomially growing
dimension $\binom{N+3}{3}$, so all seven topologies can be simulated on
the same footing without truncation up to atom numbers of at least
$N=50$. We first describe the numerical scheme itself, then use it to
(i) visualize the probability flow directly on the $\SU(4)$ tetrahedron
(\Cref{subsec:flow}), (ii) extract and compare the total intensity emitted
 $I(t)$ across all seven configurations (\Cref{subsec:intensity}),
and (iii) fit the peak-intensity scaling and inspect the level
populations (\Cref{subsec:scaling}).

We solve the population-rate equation by constructing the sparse rate
matrix \(M\) associated with \cref{eq:all-six-channels}. The probability
vector \(p(t)\), whose components are the occupation-state probabilities
\(p(q_1,q_2,q_3,q_4,t)\), obeys equation, 
${dp(t)}/{dt}=Mp(t)$.
For large \(N\), forming the full matrix exponential \(\exp(Mt)\) is
impractical. The emitted intensity is evaluated for up to $N=50$ at each time step from
\cref{eq:intensity}, while only a small number of probability snapshots are
stored for visualization on the \(\mathrm{SU}(4)\) tetrahedral lattice.

\subsection{Probability flow on the \texorpdfstring{$\SU(4)$}{SU(4)} tetrahedron}
\label{subsec:flow}

\begin{figure*}[!t]
\centering
\begin{subfigure}{0.15\textwidth}
    \includegraphics[width=\linewidth]{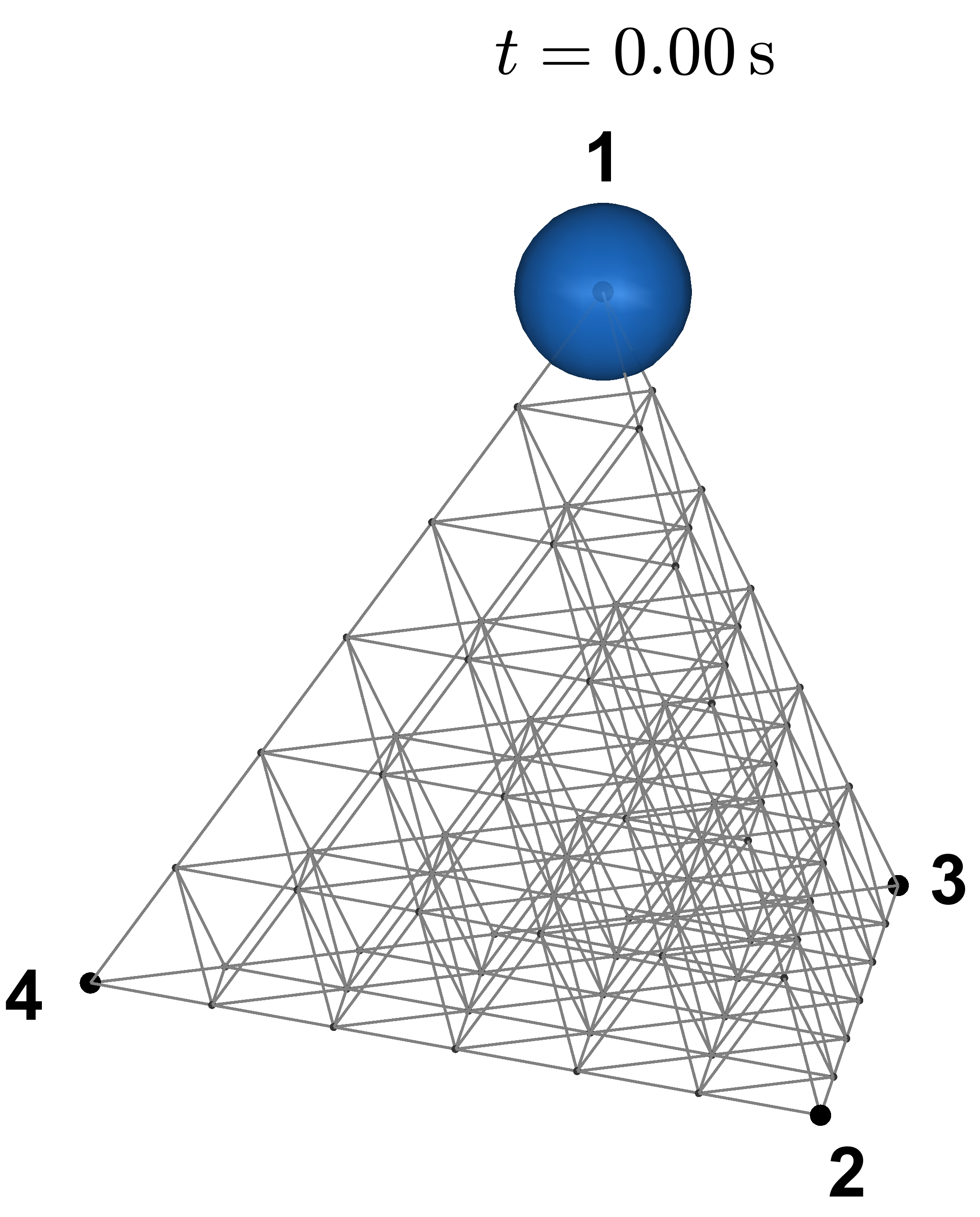}
    \caption{}
\end{subfigure}
\hfill
\begin{subfigure}{0.15\textwidth}
    \includegraphics[width=\linewidth]{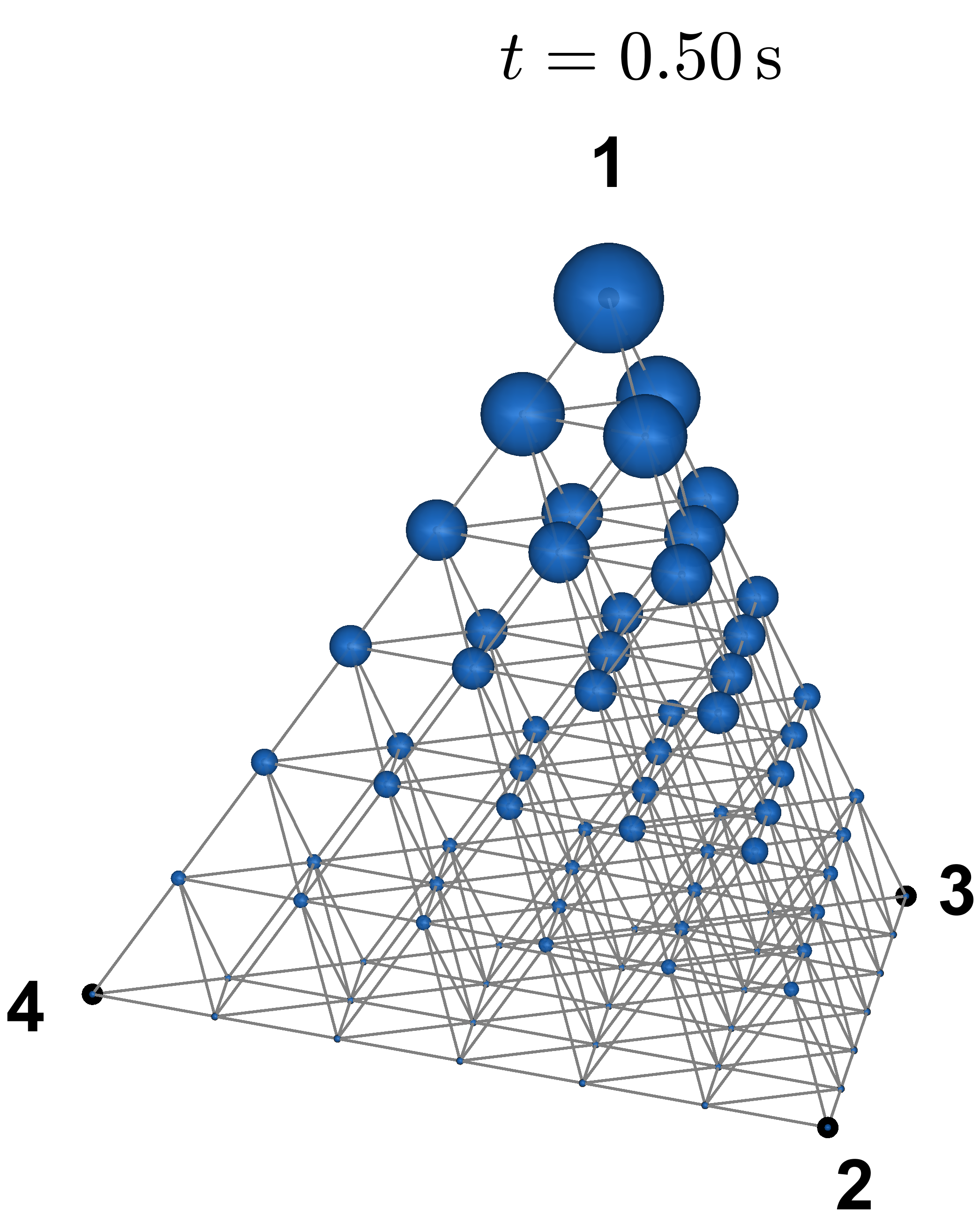}
    \caption{}
\end{subfigure}
\hfill
\begin{subfigure}{0.15\textwidth}
    \includegraphics[width=\linewidth]{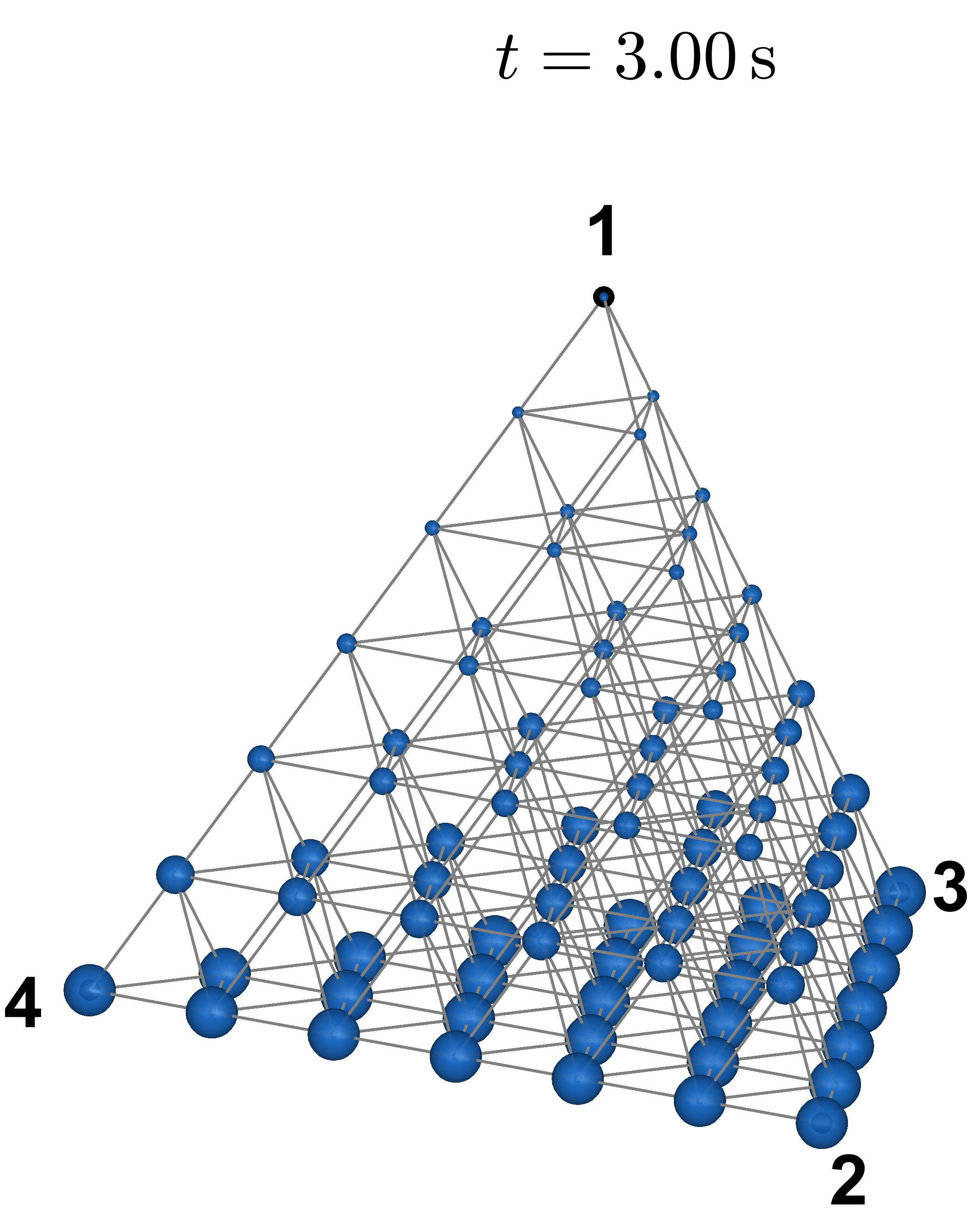}
    \caption{}
\end{subfigure}
\hfill
\begin{subfigure}{0.15\textwidth}
    \includegraphics[width=\linewidth]{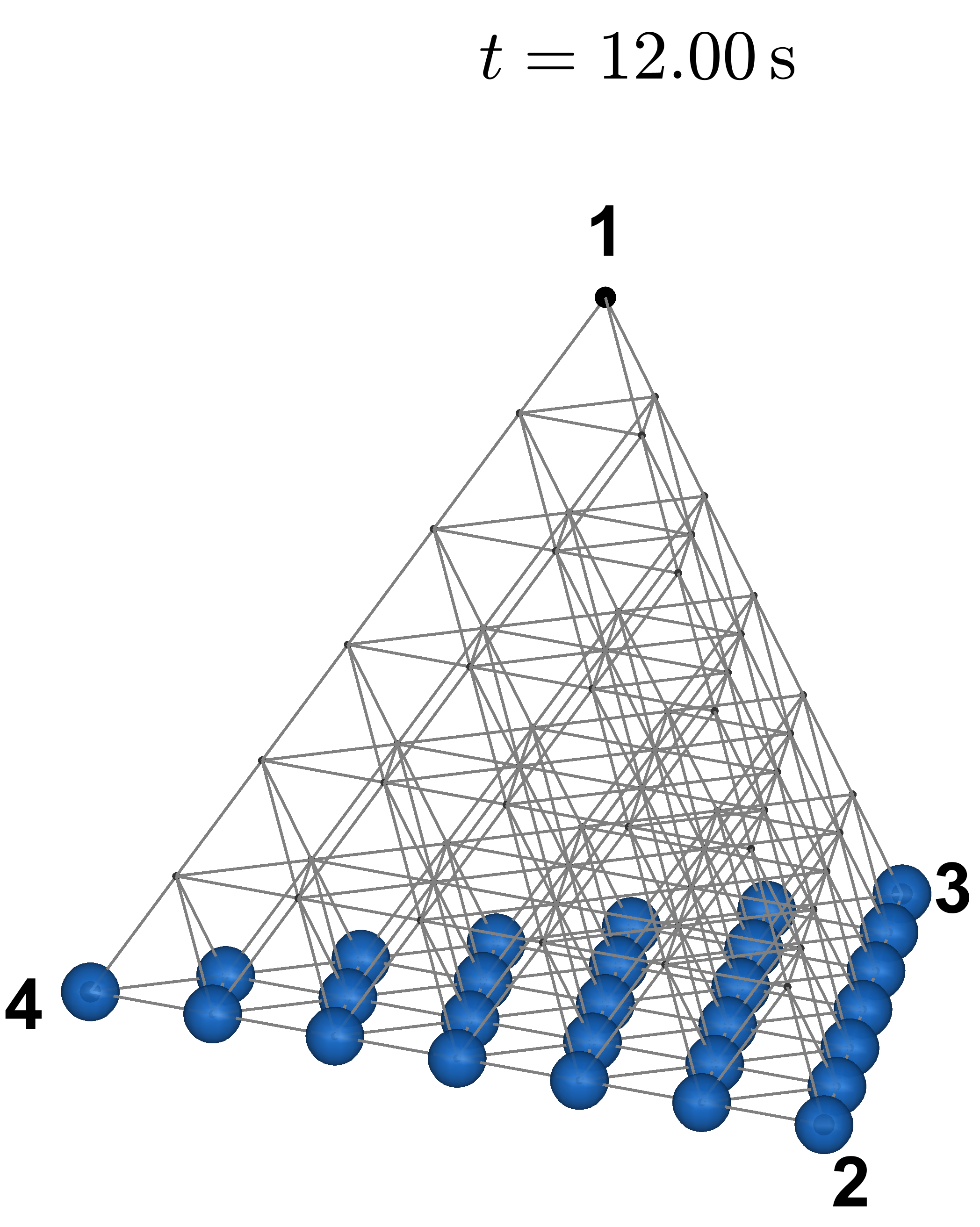}
    \caption{}
\end{subfigure}

\caption{Tripod configuration for $N=6$ with the initial condition
$p(6,0,0,0;t=0)=1$ at the top vertex, panel~(a). During the evolution, the
probability flows along the three allowed transitions $1\to2$, $1\to3$, and
$1\to4$ toward the lower face of the tetrahedron. At late times, it remains on
the $q_1=0$ plane because level 1 has emptied and no further decay is possible
in the tripod scheme.}
\label{fig:tripod-flow}
\end{figure*}

The most direct way to visualize the dynamics generated by \cref{eq:pauli-rate-compact}
is to follow the probability $p(q,t)$ over the $\SU(4)$ weight lattice as a
function of time. Each site of \cref{fig:tripod-flow} carries a ball whose
volume is proportional to $p$ at that site. The accompanying MATLAB script
generates the corresponding animation interactively; here we show two
representative time strips for $N=6$.

\begin{figure*}[!t]
\centering
\begin{subfigure}{0.15\textwidth}
    \includegraphics[width=\linewidth]{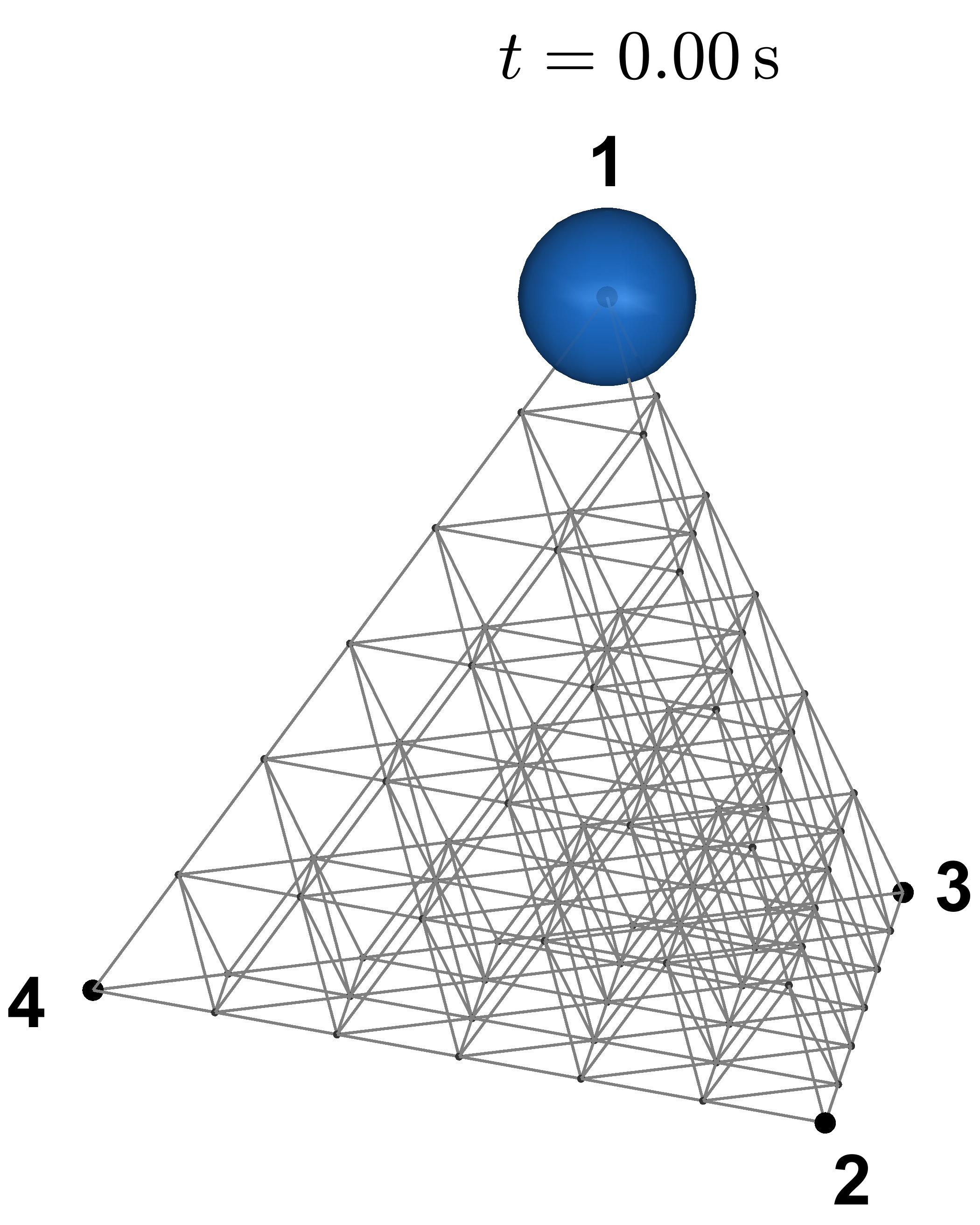}
    \caption{}
\end{subfigure}
\hfill
\begin{subfigure}{0.15\textwidth}
    \includegraphics[width=\linewidth]{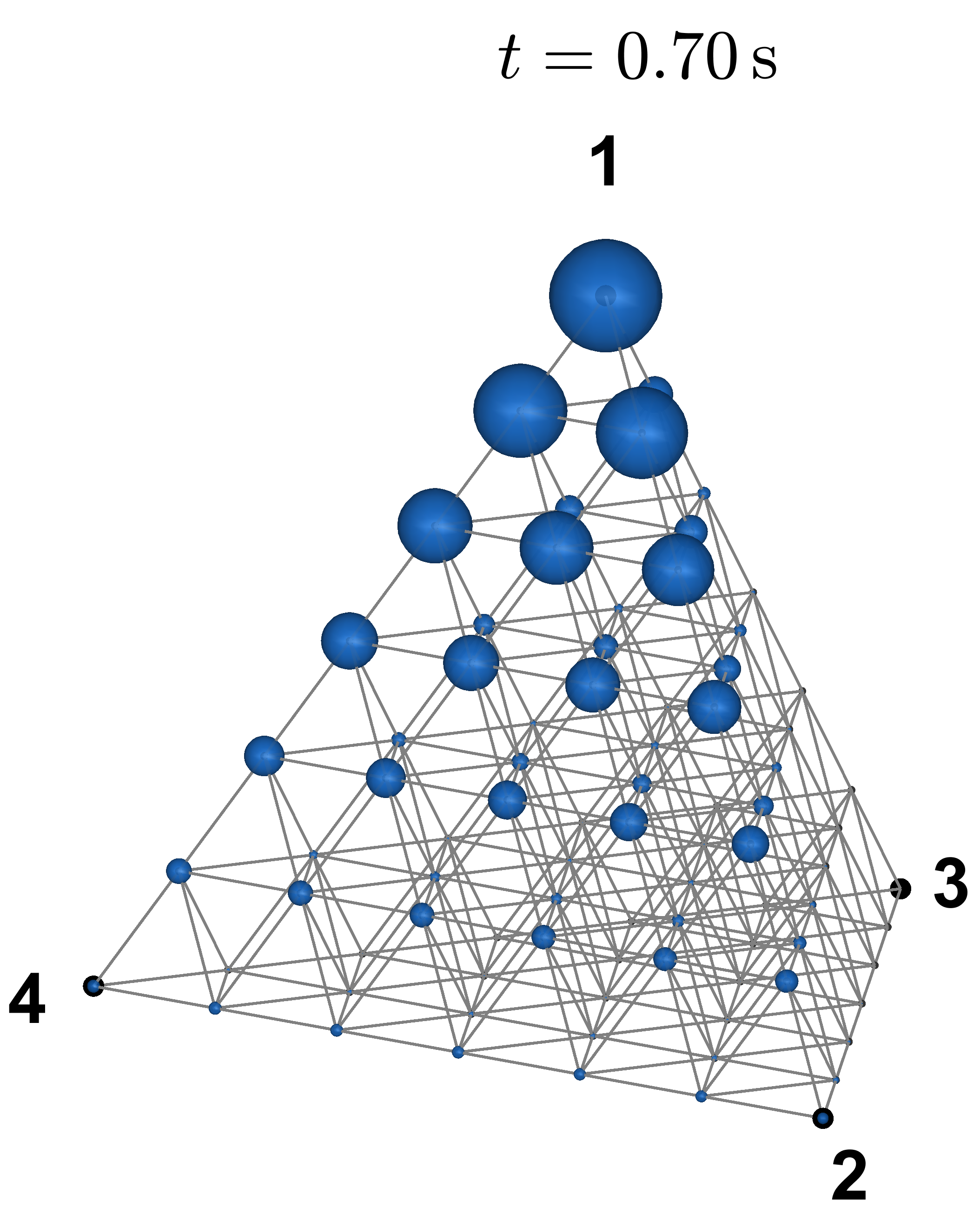}
    \caption{}
\end{subfigure}
\hfill
\begin{subfigure}{0.15\textwidth}
    \includegraphics[width=\linewidth]{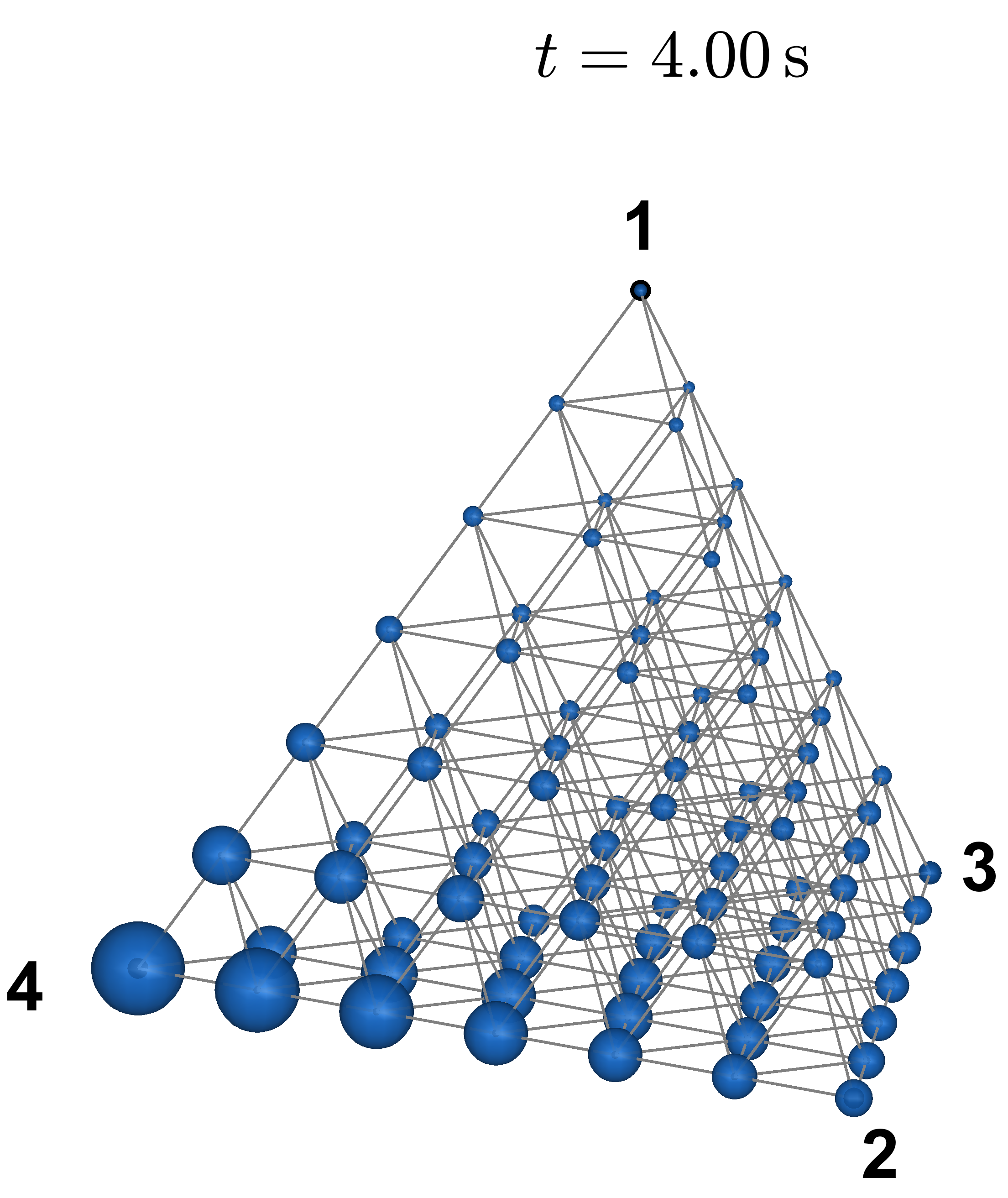}
    \caption{}
\end{subfigure}
 \hfill
\begin{subfigure}{0.15\textwidth}
    \includegraphics[width=\linewidth]{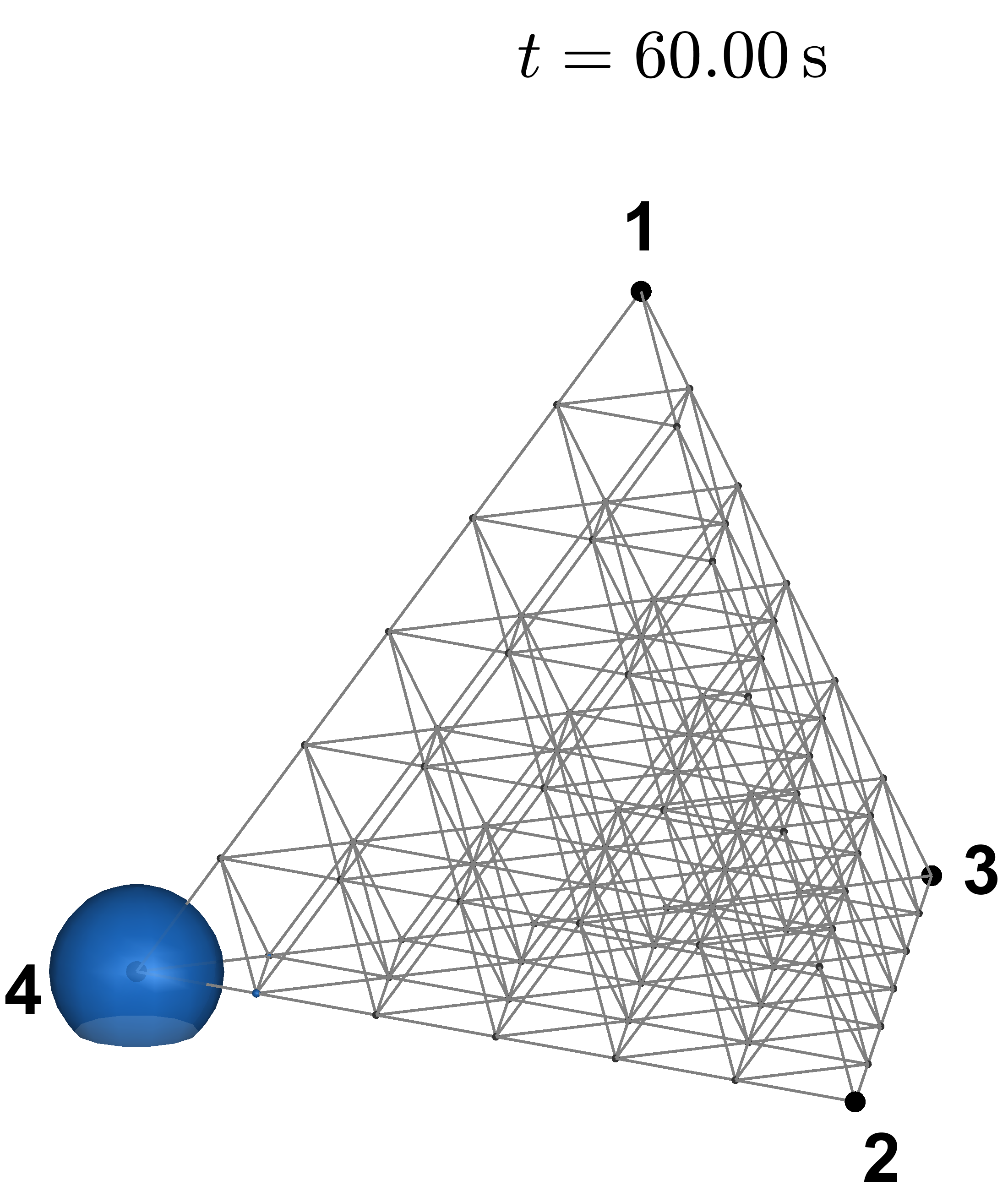}
    \caption{}
\end{subfigure}

\caption{Closed-cascade configuration for $N=6$ with all probability initially
placed at the top vertex, $p(6,0,0,0;t=0)=1$, panel~(a). The probability flows
through both the stepwise cascade $1\to2\to3\to4$ and the direct transition
$1\to4$. As time increases, population is transferred toward level 4, and the
distribution finally concentrates near the $q_4=N$ vertex.}
\label{fig:closed-cascade-flow}
\end{figure*}

\begin{figure*}[!t]
\centering
\begin{subfigure}{0.15\textwidth}
    \includegraphics[width=\linewidth]{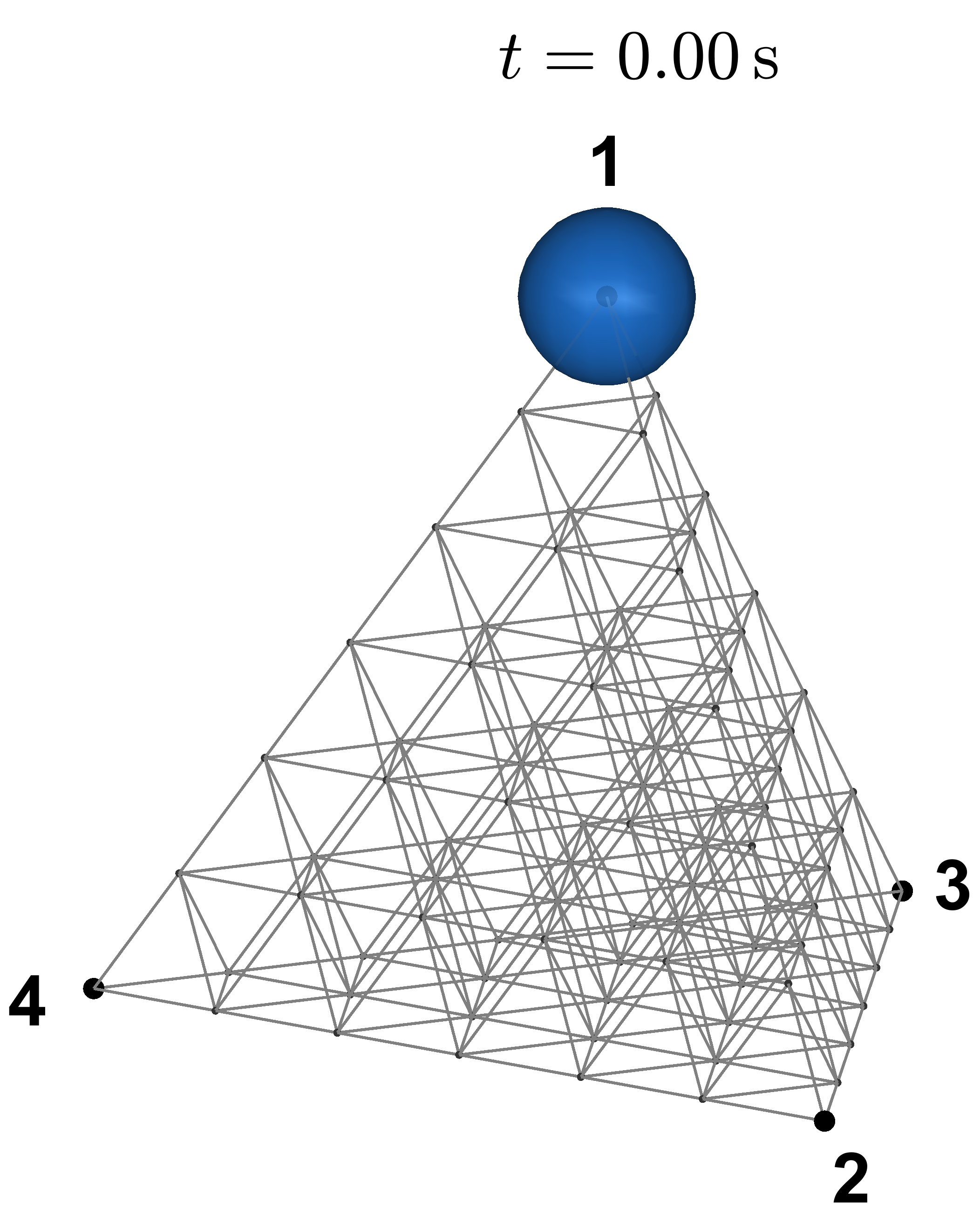}
    \caption{}
\end{subfigure}
\hfill
\begin{subfigure}{0.15\textwidth}
    \includegraphics[width=\linewidth]{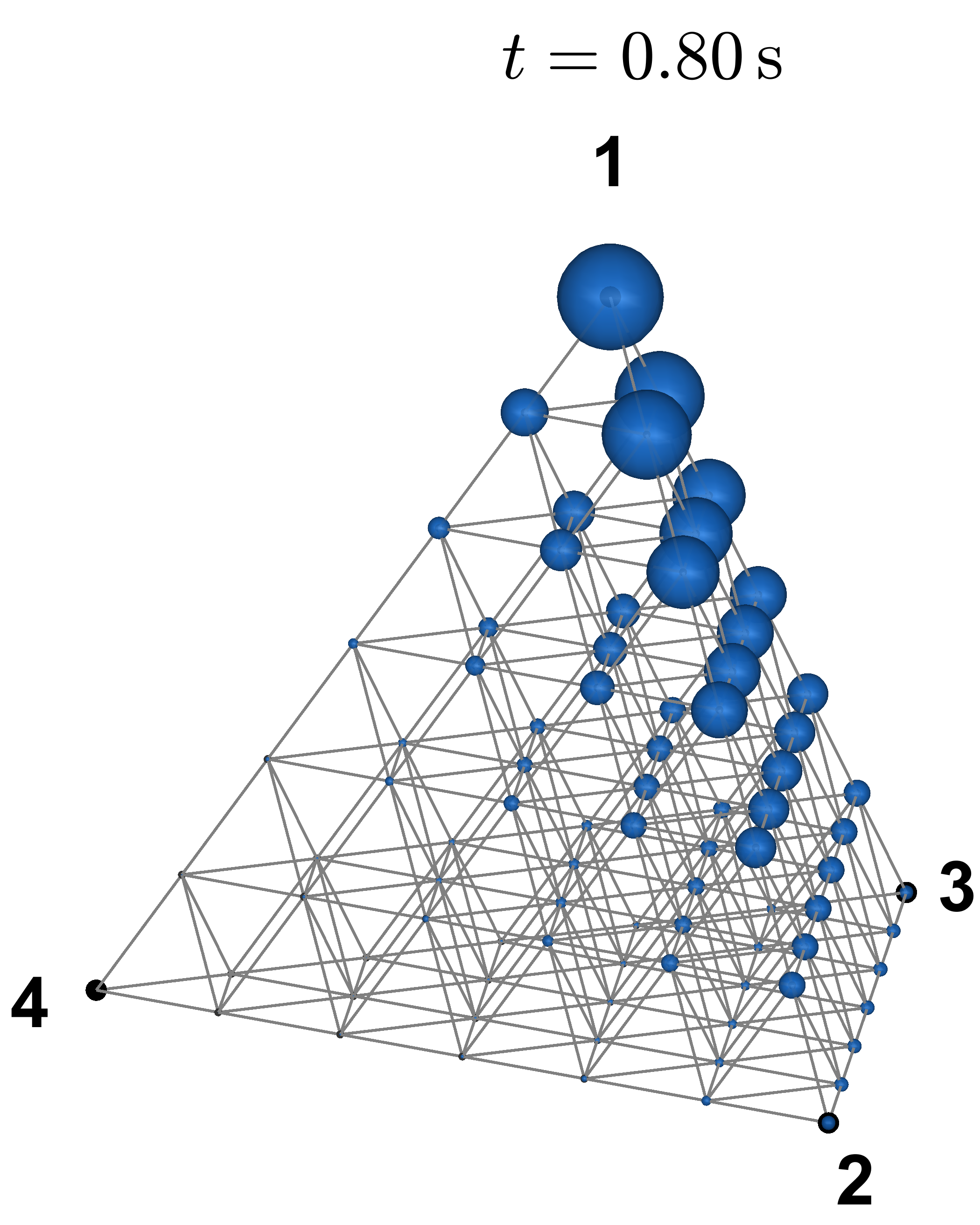}
    \caption{}
\end{subfigure}
\hfill
\begin{subfigure}{0.15\textwidth}
    \includegraphics[width=\linewidth]{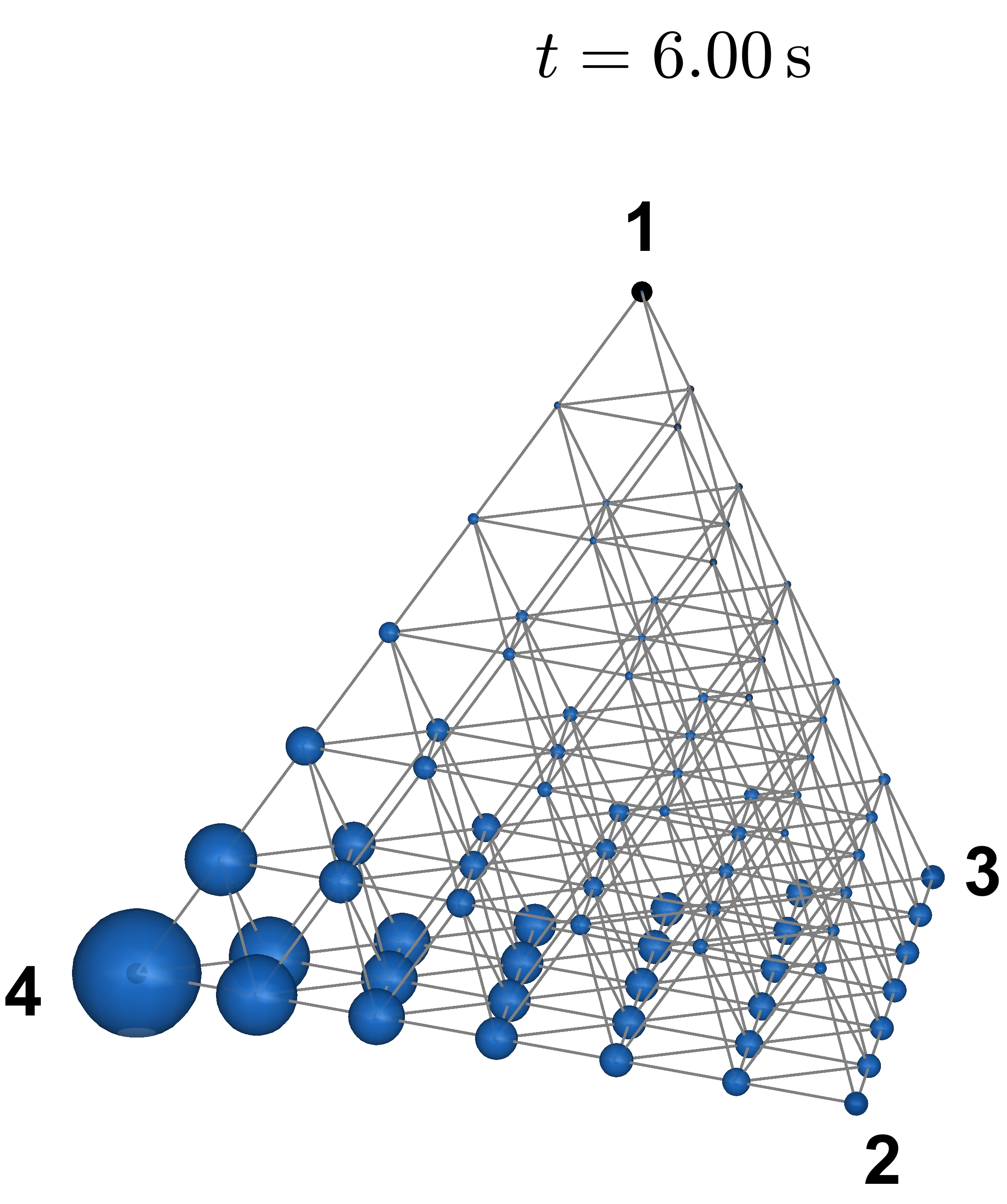}
    \caption{}
\end{subfigure}
\hfill
\begin{subfigure}{0.15\textwidth}
    \includegraphics[width=\linewidth]{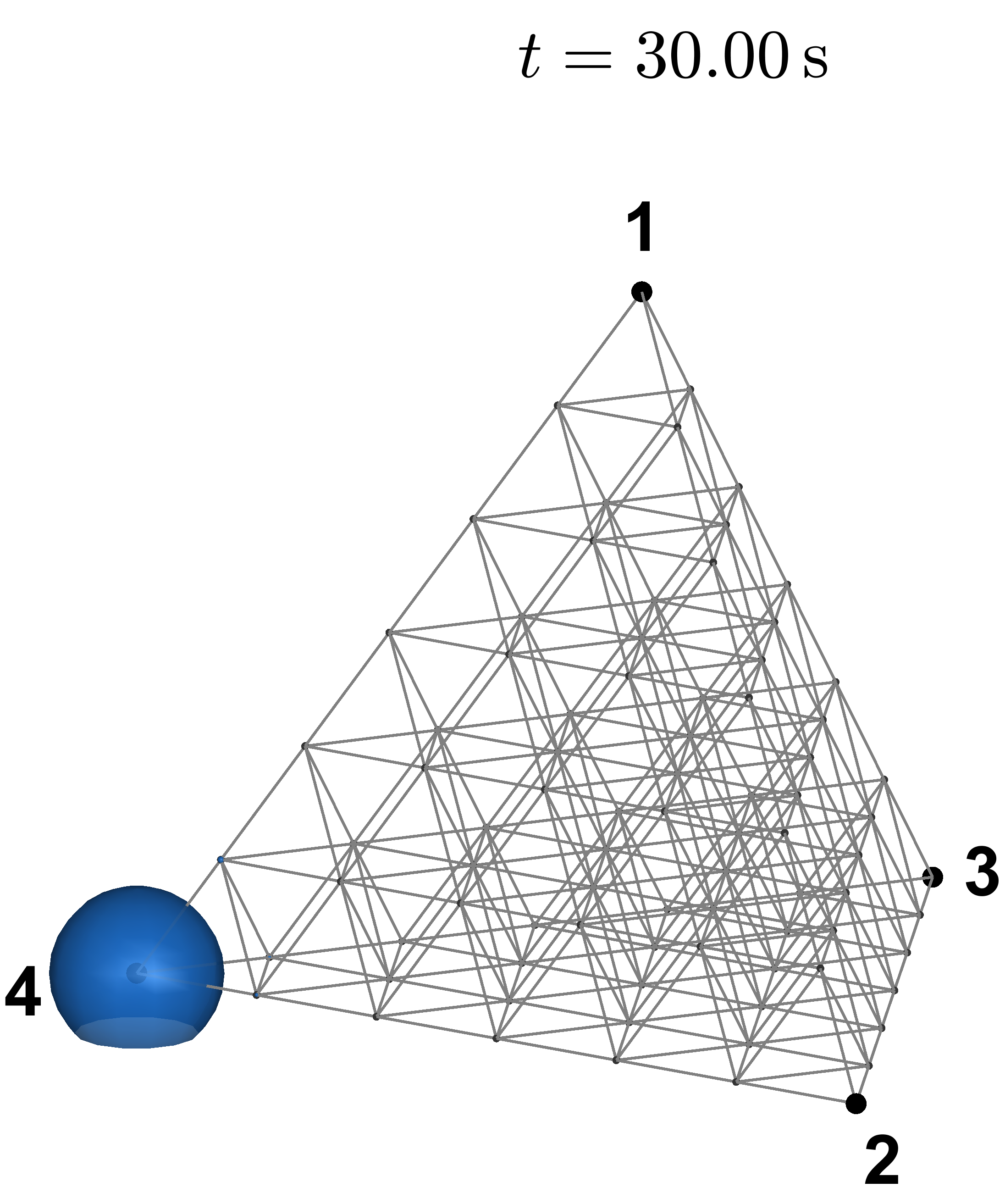}
    \caption{}
\end{subfigure}

\caption{Diamond configuration for $N=6$ with all probability initially placed
at the top vertex, $p(6,0,0,0;t=0)=1$, panel~(a). The probability first flows
from level 1 into the two intermediate levels through the transitions
$1\to2$ and $1\to3$. It then continues to the lowest level through
$2\to4$ and $3\to4$. At late times, the population accumulates at level 4, so
the distribution concentrates near the $q_4=N$ vertex of the tetrahedron.}
\label{fig:diamond-flow}
\end{figure*}
The contrast between the three configurations shown in
\cref{fig:tripod-flow,fig:closed-cascade-flow,fig:diamond-flow} is
informative. In the tripod, the three decay channels from level~1 are
completely decoupled: each atom falls independently into one of the
three lower levels, so the asymptotic distribution is trinomial across
the base of the tetrahedron and never leaves the $q_1=0$ plane. In the
closed cascade, in contrast, the sequential pathway
$1\!\to\!2\!\to\!3\!\to\!4$ shuttles probability along a single edge of
the tetrahedron while the direct channel $1\!\to\!4$ opens a shortcut,
so the flow is one-dimensional in character but split between two
routes. In the diamond, the two symmetric intermediate branches
$1\!\to\!\{2,3\}\!\to\!4$ collect the probability into a single dark
vertex $\ket{0,0,0,N}$, producing a clean funneling pattern in the
weight diagram. These three qualitatively different flow patterns
provide the visual and dynamical signatures on which the
configuration-dependent emission profiles of the next subsection can be
read off directly.
\FloatBarrier

\subsection{Total emitted intensity}
\label{subsec:intensity}

\begin{figure*}[!t]
\centering
\includegraphics[width=0.85\linewidth]{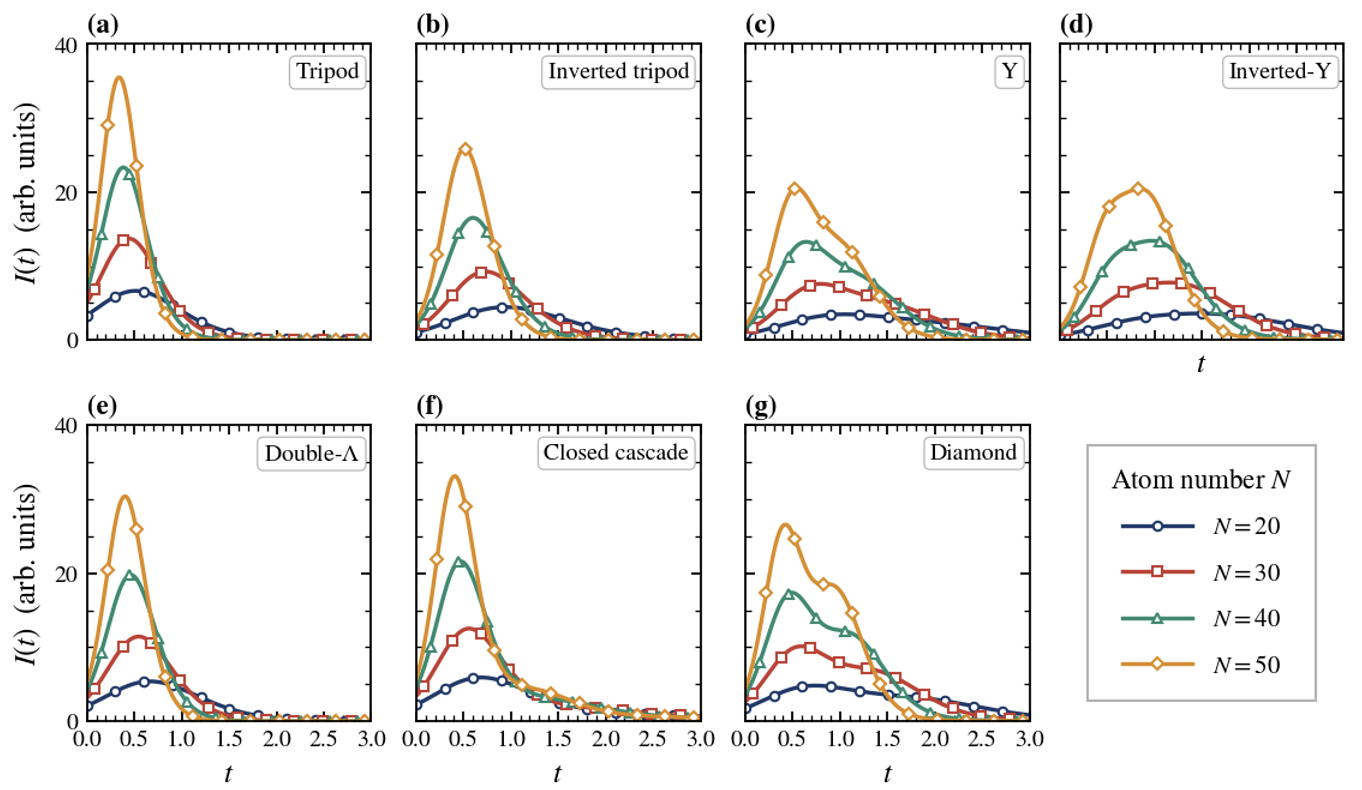}
\caption{\textbf{Collective emission dynamics for the seven four-level
topologies.} Total emitted intensity $I(t)$ for the (a) tripod, (b) inverted
tripod, (c) Y, (d) inverted-Y, (e) double-$\Lambda$, (f) closed cascade, and
(g) diamond configurations. The curves correspond to atom numbers
$N=20$, $30$, $40$, and $50$. The initial conditions are
$P_{\mathrm{Tripod}}(N,0,0,0;t=0)=1$,
$P_{\mathrm{Inv.\ tripod}}(N-2\lfloor N/3\rfloor,\lfloor N/3\rfloor,\lfloor N/3\rfloor,0;t=0)=1$,
$P_{\mathrm{Y}}(N-\lfloor N/2\rfloor,\lfloor N/2\rfloor,0,0;t=0)=1$,
$P_{\mathrm{Inv.\ Y}}(N,0,0,0;t=0)=1$,
$P_{\mathrm{Double}\text{-}\Lambda}(N-\lfloor N/2\rfloor,\lfloor N/2\rfloor,0,0;t=0)=1$,
$P_{\mathrm{Closed\ cascade}}(N,0,0,0;t=0)=1$, and
$P_{\mathrm{Diamond}}(N,0,0,0;t=0)=1$.
All other occupation states have zero initial probability.
All decay rates are equal, $\gamma_{nm}=\gamma=0.08$, and the transition
weights are $\omega_{21}=0.20$, $\omega_{31}=0.35$, $\omega_{41}=0.50$,
$\omega_{32}=0.15$, $\omega_{42}=0.30$, and $\omega_{43}=0.15$ in arbitrary
units. All configurations exhibit a delayed cooperative burst whose peak
increases quadratically with $N$. The diamond configuration shows a two-peak
profile associated with the sequential $1\rightarrow\{2,3\}\rightarrow4$
pathway, while the closed cascade develops a sharp initial burst followed by a
long decay tail due to the competing direct $1\rightarrow4$ transition. The
dynamics are obtained from the exact propagator of the population master
equation in the permutation-symmetric basis.}
\label{fig:intensity}
\end{figure*}

The tetrahedral flow patterns of \Cref{subsec:flow} translate directly
into distinct temporal profiles of the emitted radiation, because the
collective enhancement factor $q_m(q_n+1)$ in \Cref{eq:intensity} is
maximized precisely when the flow visits lattice sites where both the
emitting and the receiving level are macroscopically occupied.
\Cref{fig:intensity} shows the total emitted intensity $I(t)$ for the
seven four-level configurations at $N=20$, $30$, $40$, and $50$. For a
fixed atom number $N$, the population master equation of
\Cref{eq:all-six-channels} is solved in the permutation-symmetric
occupation-number basis $\ket{q_1,q_2,q_3,q_4}$. For each allowed transition $i\rightarrow j$, the collective decay rate from
the state $\ket{q_1,q_2,q_3,q_4}$ contains the factor $q_i(q_j+1)$, where $q_i$ is the
population of the emitting level and $(q_j+1)$ is the bosonic enhancement
factor associated with the target level. After obtaining the time-dependent
probabilities $\mathbf{P}(t)$, the total emitted intensity is evaluated from
\Cref{eq:intensity}.

For each topology, the initial condition is chosen as a single occupation state
with unit probability, while all other occupation states have zero initial
probability. The initial states used in \Cref{fig:intensity} are
\[
\begin{aligned}
\text{Tripod:}\quad
& P(N,0,0,0;t=0)=1,\\
\text{Inverted tripod:}\quad
& P\!\left(
N-2\left\lfloor\frac{N}{3}\right\rfloor,
\left\lfloor\frac{N}{3}\right\rfloor,
\left\lfloor\frac{N}{3}\right\rfloor,
0;t=0
\right)=1,\\
\text{Y:}\quad
& P\!\left(
N-\left\lfloor\frac{N}{2}\right\rfloor,
\left\lfloor\frac{N}{2}\right\rfloor,
0,0;t=0
\right)=1,\\
\text{Inverted-Y:}\quad
& P(N,0,0,0;t=0)=1,\\
\text{Double-}\Lambda\text{:}\quad
& P\!\left(
N-\left\lfloor\frac{N}{2}\right\rfloor,
\left\lfloor\frac{N}{2}\right\rfloor,
0,0;t=0
\right)=1,\\
\text{Closed cascade:}\quad
& P(N,0,0,0;t=0)=1,\\
\text{Diamond:}\quad
& P(N,0,0,0;t=0)=1 .
\end{aligned}
\]

In all configurations, the emission begins from a relatively weak initial rate
and then develops a delayed cooperative burst. At early times, the intensity is
approximately proportional to $N$, because only the initially populated
transitions contribute. During the evolution, population accumulates in lower
or intermediate levels, increasing the enhancement factor $(q_j+1)$. This
raises the emission rate above the independent-emitter scaling and produces a
collective peak that grows superlinearly with $N$.

The tripod configuration illustrates this mechanism clearly. The initial state
$\ket{N,0,0,0}$ decays through three downward transitions into different lower
levels. Although the probability flow is split among several branches, each
branch can still acquire a macroscopic occupation. Therefore the collective
enhancement is not removed by the branching process, and the tripod reaches one
of the largest peak intensities among the seven topologies.

The configurations also differ in the structure of their emission bursts. The
tripod, inverted tripod, inverted-Y, and double-$\Lambda$ cases show essentially
single dominant cooperative bursts. By contrast, the Y, diamond, and closed
cascade configurations develop two-stage intensity profiles. These two maxima
are not identical superradiant bursts repeated at different times. They are
produced by different groups of transitions that become active at different
stages of the population flow.

In the Y configuration, the first peak is associated with decay out of
the initially populated upper manifold into an intermediate level. The
second peak appears only after this intermediate level has acquired a
macroscopic population and subsequently decays to the lower state. In the
diamond configuration, the first peak corresponds mainly to the
splitting stage $1\rightarrow\{2,3\}$, while the second peak is generated
by the later relaxation stage $\{2,3\}\rightarrow4$ into the common lower
level. The closed cascade shows a less symmetric two-stage structure
because the direct transition $1\rightarrow4$ competes with the
sequential pathway $1\rightarrow2\rightarrow3\rightarrow4$.

\subsection{Peak-intensity scaling and level populations}
\label{subsec:scaling}

The temporal transients of \Cref{subsec:intensity} show that all seven
configurations exhibit a delayed cooperative burst whose peak grows
quadratically with atom number. To make this observation quantitative,
we now extract the peak height $I_{\mathrm{peak}}(N)=\max_{t}I(t)$
directly from the numerical solutions of \Cref{eq:all-six-channels} and
fit it to a topology-dependent power law. This gives a single pair of
scaling parameters $(a,p)$ per configuration and allows the seven
four-level topologies to be compared on the same axes.
\begin{figure*}[t!]
\centering
\includegraphics[width=0.75\linewidth]{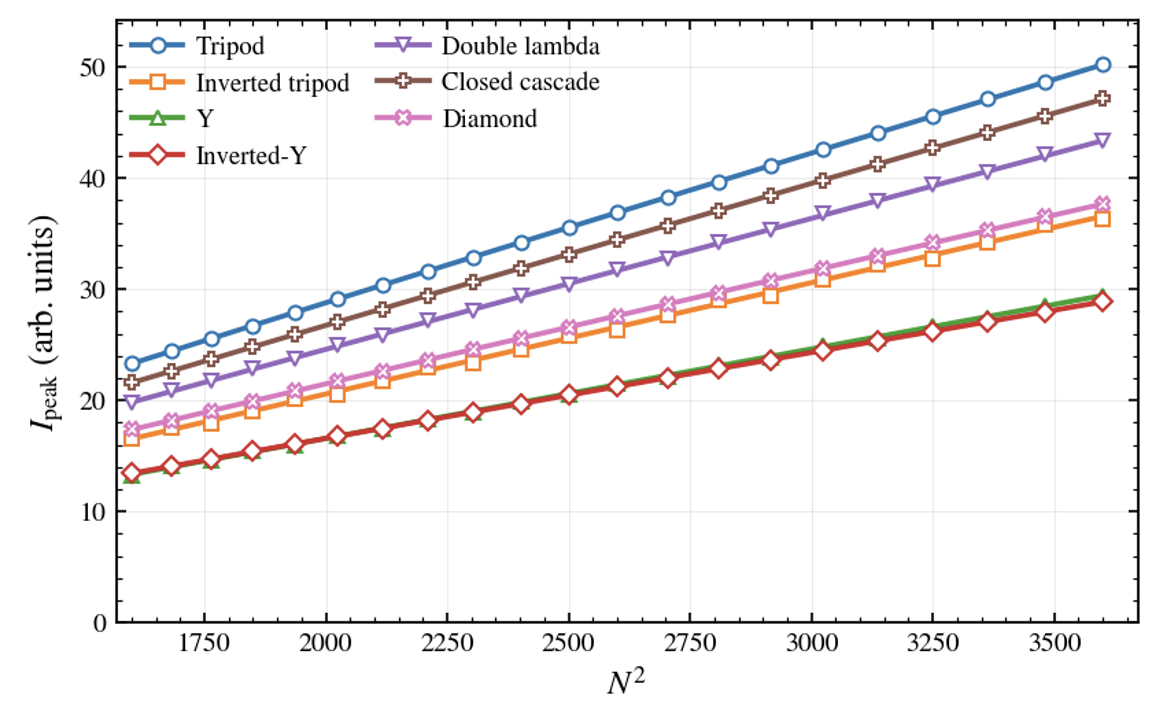}
\caption{\textbf{Power-law scaling of the peak emitted intensity.} Peak
emitted intensity $I_{\mathrm{peak}}=\max_{t}I(t)$ for the seven four-level
topologies over the range $N=20,\ldots,40$, plotted against $N^{2}$ so that
nearly quadratic cooperative growth appears approximately linear. The symbols show
the numerical results, while the solid lines are least-squares fits to
\(I_{\mathrm{peak}}=aN^p\). The seven topologies are shown on the same axes
to allow direct comparison of their absolute peak intensities and fitted
scaling exponents. The fit parameters are: tripod, $a=0.029$, $p=1.812$;
inverted tripod, $a=0.0138$, $p=1.921$; Y, $a=0.0111$, $p=1.922$;
inverted-Y, $a=0.0123$, $p=1.896$; double-$\Lambda$, $a=0.0198$, $p=1.872$;
closed cascade, $a=0.0218$, $p=1.869$; and diamond, $a=0.0188$, $p=1.851$.
All fits have $R^{2}\simeq1$, so that the finite-size power law
$I_{\mathrm{peak}}=aN^{p}$ describes the numerical data very well over the
plotted interval. The exponents are larger than unity but below the ideal
quadratic value $p=2$, indicating a superlinear but sub-Dicke cooperative
enhancement within the finite-$N$ crossover regime accessible in the
present simulations.}
\label{fig:peak-scaling}
\end{figure*}
\Cref{fig:peak-scaling} shows $I_{\mathrm{peak}}(N)$ for the seven
four-level configurations over the range $N=20,\ldots,40$. The
horizontal axis is chosen as $N^{2}$ so that a nearly quadratic
atom-number dependence would appear approximately linear in the plotted
representation. The data are not, however, fitted to a fixed $N^{2}$
law: instead, each topology is fitted independently to the power law
$I_{\mathrm{peak}}(N)=aN^{p}$. The seven overlaid curves therefore allow
both the absolute peak strengths and the best-fit exponents to be read
off simultaneously.

The fitted parameters show that all configurations have superlinear growth,
with exponents in the range $p\simeq1.81$--$1.92$. The largest peak intensity is
obtained for the tripod configuration, which has the largest scaling factor
$a=0.029$. This agrees with the intensity dynamics in \Cref{fig:intensity}: the initially
occupied upper level decays through three allowed transitions, so the total
emission receives simultaneous contributions from all three branches of the
tripod. 

The Y and inverted-tripod configurations show the largest exponents,
$p=1.922$ and $p=1.921$, respectively. Their scaling factors are smaller,
$a=0.0111$ for the Y configuration and $a=0.0138$ for the inverted tripod, so
their absolute peak intensities remain below the tripod peak. This means that
these topologies show stronger relative growth with $N$, but smaller absolute
emission strength in the plotted range. The inverted-Y configuration also has a
large exponent, $p=1.896$, with an almost perfect log--log fit
($R^2\simeq1$), indicating a very regular finite-size power-law scaling.

The double-$\Lambda$, closed cascade, and diamond configurations form an
intermediate group. Their scaling factors,
$a=0.0198$, $0.0218$, and $0.0188$, are larger than those of the Y and
inverted-tripod cases, but smaller than that of the tripod. Their exponents,
$p=1.872$, $1.869$, and $1.851$, also lie between the tripod and the
Y/inverted-tripod values. Thus, the comparison separates the topologies into
two effects: the scaling factor $a$ controls the absolute peak strength, while the
exponent $p$ measures how rapidly the peak grows with atom number over the
finite interval considered. All values of $R^2$ are close to unity, confirming
that the power-law form $I_{\mathrm{peak}}=aN^p$ accurately describes the
numerical data for $N=20$--$40$.

\begin{figure*}[t!]
\centering
\includegraphics[width=0.85\linewidth]{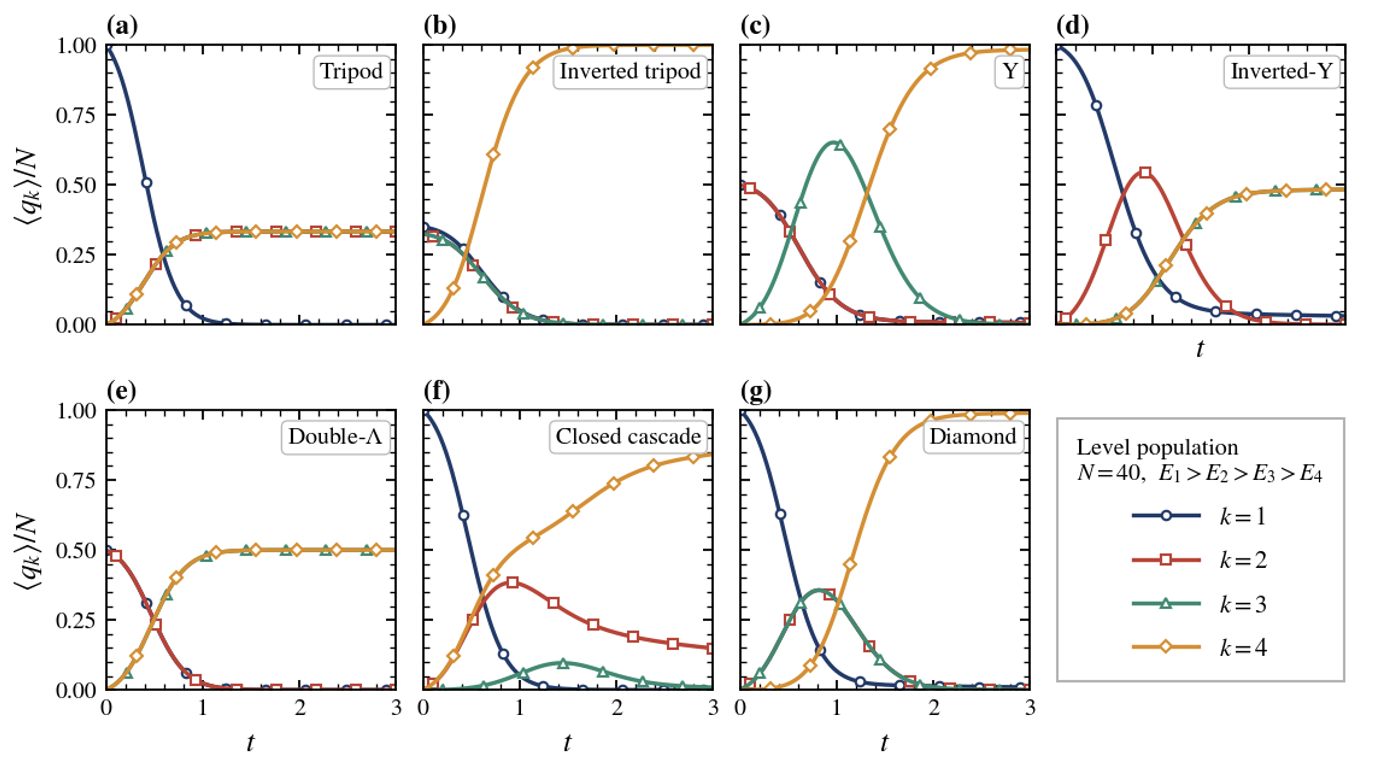}
\caption{\textbf{Level-population dynamics at $N=40$.} Normalized mean
populations $\langle q_k(t)\rangle/N$ of levels $k=1,2,3,4$ for the seven
four-level topologies, with energy ordering $E_1>E_2>E_3>E_4$. The panel
ordering is the same as in \Cref{fig:intensity}: (a) tripod, (b) inverted
tripod, (c) Y, (d) inverted-Y, (e) double-$\Lambda$, (f) closed cascade, and
(g) diamond. The same initial states and transition parameters are used as in
the total-emission calculation. The tripod redistributes the initial population
among three lower levels, whereas the inverted tripod, Y, and diamond funnel
the population into the common lowest level~4. The inverted-Y and
double-$\Lambda$ terminate in two-level mixtures. The closed cascade displays a
slow relaxation tail because direct and sequential decay pathways compete.}
\label{fig:populations}
\end{figure*}

\Cref{fig:populations} shows the normalized mean level populations
$\langle q_k(t)\rangle/N$ for $N=40$, using the same initial states and the same
transition parameters as in the total-intensity calculation. These population
curves explain the differences between the intensity profiles and the fitted
peak-scaling parameters.

In the tripod configuration, the initial population in level~1 decays
into levels 2, 3, and 4. For the equal decay rates used here,
$\gamma_{21}=\gamma_{31}=\gamma_{41}$, the three lower levels are
dynamically equivalent, so each of the three branches acquires a
macroscopic occupation of order $N/3$. Every branch therefore
contributes to the collective enhancement factor $q_1(q_j+1)$ in
\Cref{eq:intensity}, and the tripod attains one of the largest absolute
peak intensities in \Cref{fig:peak-scaling}. If the three rates were
instead taken unequal, this three-fold degeneracy would be lifted and
the branch with the largest $\gamma$ would eventually dominate; the
transition frequencies $\omega_{nm}$, by contrast, do not affect the
population flow in this rate equation at all---they only enter the
emitted intensity as photon-energy weights, so tuning them changes
$I_{\mathrm{peak}}$ without altering the level dynamics.

The inverted tripod, Y, and diamond configurations act as funnel-like
systems. Their populations eventually accumulate in the common lowest
level~4, but they arrive there by qualitatively different routes. The
Y and diamond panels show a clear intermediate population build-up
before the final relaxation into level~4. This intermediate build-up is
the microscopic origin of the two-stage intensity profile seen for
these two configurations in \Cref{fig:intensity}: the first emission
maximum is associated with decay from the initially occupied upper
levels into the intermediate manifold, while the second appears once
the intermediate population itself decays further into the terminal
lower state.

The inverted-Y and double-$\Lambda$ configurations end in two-level mixtures
rather than in a single final state. In the inverted-Y case, the initial
population in level~1 first feeds level~2 and then separates between levels 3
and 4. In the double-$\Lambda$ case, the initially occupied levels 1 and 2 decay
into levels 3 and 4, producing an approximately balanced final distribution.
Finally, the closed cascade shows the slowest relaxation among the displayed
cases. A visible fraction of the population remains in level~2 over the plotted
time window, while level~4 continues to grow. This slow relaxation is caused by
the competition between the direct transition $1\rightarrow4$ and the
sequential pathway $1\rightarrow2\rightarrow3\rightarrow4$.

\section{Discussion}
\label{sec:discussion}

The $\SU(4)$-symmetric master equation derived above accomplishes three things
in a single framework. First, it unifies all seven dipole-allowed four-level
configurations under one Pauli-type rate equation whose only
configuration-dependent input is the set of allowed channels. In each special case,
our results agree with the configuration-specific master equations that have
been derived previously \cite{mazets2005,niu2002,cho2016,tey2008}, but
unlike those derivations, our approach makes the underlying $\SU(4)$ symmetry
explicit and produces the master equation by purely algebraic means. Second, the
Bose-enhanced rate factor $q_m(q_n+1)$ in \cref{eq:pauli-rate-compact} generalizes
Dicke's $(J+M)(J-M+1)$ factor and makes the geometric origin of
superradiance---namely the representation theory of the symmetric group on $N$
atoms---manifest at every level of the cascade. Third, the formalism reduces an
in-principle exponentially large many-body problem to a polynomial-sized linear
ODE on the lattice of fully symmetric $\SU(4)$ states, with $O(N^3)$ equations
rather than $O(4^N)$, which makes it numerically very efficient.

Several aspects of the present treatment can be relaxed or extended. (i) The
diagonal Pauli-type rate equation neglects atom--atom coherence in the
symmetric subspace. For many phenomena, such as dissipative
cooling, and photon-emission statistics, the coherence contribute only
subleading corrections to the leading $N$-scaling, but their inclusion is needed
to capture the full biphoton statistics in cascade and double-$\Lambda$
geometries \cite{wang2015,cho2016}. This is a promising direction for future
$\SU(4)$-based work. (ii) Inhomogeneous driving and atom--atom dipole--dipole
interactions beyond the all-to-all approximation break the permutation symmetry
exploited here. Recent work has shown that Dicke superradiance in the dilute
regime requires long-range, beyond-nearest-neighbour interactions
\cite{rubies2023}; the corresponding extension of our $\SU(4)$ formalism would
map onto a finite-range tight-binding problem on the tetrahedral lattice. (iii)
Coherent driving terms, or Rabi couplings on selected channels, can be added to
\cref{eq:lindblad-su4} and lead, within the same symmetric subspace, to
optical-Bloch equations on the $\SU(4)$ lattice---a natural framework for
double-EIT, slow light, and the photon-pair sources discussed in the
introduction.

On the experimental side, our results place the seven four-level topologies
on a common quantitative footing and identify which configurations are most
promising for observing textbook superradiant bursts at modest atom numbers.
Over the range $N=20$--$40$ analyzed here, all seven configurations exhibit
superlinear cooperative growth with fitted exponents concentrated in the
narrow window $1.81\lesssim p\lesssim 1.92$, i.e.\ superlinear. Within this window, however, two subgroups emerge: the inverted
tripod, Y, and inverted Y---topologies in which the emitted photons are
funneled into a single common lower level---attain the largest exponents
($p\simeq 1.90$--$1.92$), whereas the tripod and the branched
double-$\Lambda$, closed-cascade, and diamond schemes cluster near
$p\simeq 1.85$--$1.87$ and combine somewhat smaller exponents with the
largest absolute scaling factors $a$. This qualitative pattern is the
$\SU(4)$ analogue of the pure Dicke superradiance $N^{2}$ scaling recovered in the
three-level $\SU(3)$ trilogy of
Refs.~\cite{ariunbold2022csr,ariunbold2025lambda,ariunbold2026vee}, so a
natural first experimental target is a rare-earth or alkali cascade whose
lowest level is a common ``dark'' storage state. The closed-cascade
configuration, which contains the sequential
$1\!\to\!2\!\to\!3\!\to\!4$ pathway together with the direct
$1\!\to\!4$ channel, has an intermediate two-stage intensity profile that
maps naturally onto the incoherently pumped cascade superfluorescence in
Er:YLF \cite{chiossi2021} and onto the ultrafast cesium and rubidium
vapor experiments of Refs.~\cite{ariunbold2010,ariunbold2022,
ariunbold2022csbeats,thompson2014,ariunbold2014apl,ariunbold2012ol,
ariunbold2012pra}. On the solid-state side, hybrid perovskite
superfluorescence \cite{raino2018,findik2021,biliroglu2022,huang2022} and
diamond-nanocrystal superradiance \cite{bradac2017} occur in materials
whose low-lying multiplet structure can plausibly be modeled as a
four-level $\SU(4)$ tetrahedron, so the topology-dependent scaling
exponents reported here provide a quantitative benchmark against which
future room-temperature superradiant experiments can be compared.
Finally, the sub-photon steady-state superradiant laser \cite{bohnet2012}
and the mHz-linewidth strontium clock realization \cite{norcia2016}
suggest an experimental setting in which the collective four-level
dynamics on the $\SU(4)$ tetrahedron could be probed inside a
high-finesse cavity.

\section{Conclusion}
\label{sec:conclusion}

We have given an $\SU(4)$-symmetric extension of Agarwal's multi-level
spontaneous-emission framework to $N$ identical four-level atoms. The
fully symmetric representation of $\SU(4)$ provides the natural many-body
Hilbert space, with collective basis states $\ket{q_1,q_2,q_3,q_4}$
sitting on a tetrahedral lattice whose dimension $\binom{N+3}{3}$ grows
only as $O(N^{3})$ with the number of atoms. The action of the $\SU(4)$
generators on this basis, combined with the dissipative Lindblad
equation, collapses into a Pauli-type rate equation
[\cref{eq:pauli-rate-compact}] together with a closed-form expression for
the total emitted intensity [\cref{eq:intensity}]. We then specialized
this common master equation to the seven dipole-allowed four-level
topologies of physical interest, and solved each numerically for atom
numbers up to $N=50$. In every case, the emitted intensity develops a
delayed cooperative burst whose peak height obeys a power law
$I_{\mathrm{peak}}=aN^{p}$ over the fitted range $N=20$--$40$, with
exponents $p$ concentrated in the narrow interval $1.81\lesssim p\lesssim
1.92$ but with topology-dependent scaling factors $a$ that span more than
half a decade. Configurations whose channels funnel probability into a
common lower vertex (inverted tripod, Y, inverted Y) attain the largest
exponents, whereas the branched tripod, double-$\Lambda$, closed-cascade
and diamond schemes combine slightly smaller exponents with the largest
absolute peak intensities. The probability flow over the $\SU(4)$
tetrahedron provides a uniform, geometric visualization of all seven
configurations, and the resulting framework lays a natural foundation for
future $\SU(4)$-based studies of multi-level coherence, biphoton
correlations, and experimentally relevant superradiant systems in
four-level atomic media, as well as a direct four-level companion to our
earlier three-level $\SU(3)$
trilogy~\cite{ariunbold2022csr,ariunbold2025lambda,ariunbold2026vee}.

\section*{CRediT authorship contribution statement}
\textbf{Mend-Amar Lutsukh:} Methodology, Software, Formal analysis,
Visualization, Writing -- original draft.
\textbf{Munkh-Uchral Bazarsan:} Methodology, Software, Formal analysis,
Visualization.
\textbf{Tuguldur Begzjav:} Formal analysis, Investigation, Writing --
review \& editing.
\textbf{Gombojav O. Ariunbold:} Conceptualization, Methodology,
Supervision, Project administration, Writing -- original draft,
Writing -- review \& editing.

\section*{Declaration of competing interest}
The authors declare that they have no known competing financial
interests or personal relationships that could have appeared to
influence the work reported in this paper.

\section*{Data availability}
No experimental data were generated for this work. The numerical code
that reproduces the figures of \Cref{sec:numerics} is available from
the corresponding author upon reasonable request.

\section*{Acknowledgments}
The authors thank Professor G.~S.~Agarwal for stimulating discussions
on multi-level collective spontaneous emission. 

\printcredits

\bibliographystyle{elsarticle-num}
\bibliography{Bibliography}

\end{document}